\documentclass[prb,aps,superscriptaddress,twocolumn,amssymb,showpacs]{revtex4}

\usepackage{epsfig}
\usepackage{dcolumn}
\usepackage{bm}

\begin{document}

\title{       Orbital polarons versus itinerant $e_g$ electrons 
              in doped manganites }

\author {     Maria Daghofer }
\affiliation{ Institute of Theoretical and Computational Physics,
              Graz University of Technology, 
              Petersgasse 16, A-8010 Graz, Austria }

\author {     Andrzej M. Ole\'{s} }
\affiliation{ Marian Smoluchowski Institute of Physics, Jagellonian 
              University, Reymonta 4, PL-30059 Krak\'ow, Poland }

\author {     Wolfgang von der Linden }
\affiliation{ Institute of Theoretical and Computational Physics, 
              Graz University of Technology, 
              Petersgasse 16, A-8010 Graz, Austria }

\date{\today}

\begin{abstract}
We study an effective one-dimensional (1D) orbital $t$-$J$ model 
derived for strongly correlated $e_g$ electrons in doped manganites. 
The ferromagnetic spin order at half filling is supported by orbital 
superexchange $\propto J$ which stabilizes orbital order with 
alternating $x^2-y^2$ and $3z^2-r^2$ orbitals. In a doped system it
competes with the kinetic energy $\propto t$. When a single hole is 
doped to a half-filled chain, its motion is hindered and a localized 
{\it orbital polaron\/} is formed. An increasing doping generates 
either separated polarons or phase separation into hole-rich and 
hole-poor regions, and eventually polarizes the orbitals and gives a 
{\it metallic phase\/} with occupied $3z^2-r^2$ orbitals. This 
crossover, investigated by exact diagonalization at zero temperature, 
is demonstrated both by the behavior of correlation functions and by 
spectral properties, showing that the orbital chain with Ising 
superexchange is more classical and thus radically different from the 
1D spin $t$-$J$ model. At finite temperature we derive and investigate 
an effective 1D orbital model using a combination of exact 
diagonalization with classical Monte-Carlo for spin correlations. 
A competition between the antiferromagnetic and ferromagnetic spin 
order was established at half filling, and localized polarons were 
found for antiferromagnetic interactions at low hole doping. Finally, 
we clarify that the Jahn-Teller alternating potential stabilizes the 
orbital order with staggered orbitals, inducing the ferromagnetic spin 
order and enhancing the localized features in the excitation spectra. 
Implications of these findings for colossal magnetoresistance 
manganites are discussed.

[Published in Phys. Rev. B {\bf 70}, 184430 (2004)]
\end{abstract}


\maketitle

\section{Introduction}
\label{sec:intro}

Doped perovskite manganese oxides R$_{1-x}$A$_x$MnO$_3$, where R and 
A are rare earth and alkaline earth ions, have attracted increasing 
attention because they show a rich variety of electronic, magnetic and
structural phenomena, and several different types of ordered phases.
\cite{Dag02} To explain the colossal magnetoresistence\cite{Sch95} 
(CMR) and metal-insulator transition observed in these compounds as a 
function of either doping $x$ or temperature $T$, which suggests their 
potential technological applications, one has to go beyond the simple 
double exchange model of Zener,\cite{Zen51} and investigate a complex 
interplay between magnetic, orbital, and lattice degrees of freedom, 
as well as the conditions for the itinerant behavior of strongly 
correlated $e_g$ electrons.

The theoretical challenge is to understand the properties of doped 
manganites in terms of the dynamics of correlated $e_g$ electrons 
which involves their orbital degrees of freedom.\cite{Dag02,Ram04} 
Although several features such as the transition to ferromagnetic 
(FM) order under doping, and the magnetic excitations in the FM phase, 
were qualitatively reproduced within the Kondo model which assumes a 
nondegenerate conduction band,
\cite{Dag02,Dag98,Fur98,Edw99,Ali01,San02,Kol02,Kol03} it is evident 
that the degeneracy of $e_g$ orbitals plays a major role in the double 
exchange model, in the transport properties of doped manganites, and in 
the CMR effect itself.\cite{Hor99,Bri99,Ole02} In the undoped LaMnO$_3$ 
charge fluctuations are suppressed by large on-site Coulomb interaction 
$U$, leading to superexchange which involves both spin and orbital $e_g$ 
degrees of freedom. 

Purely electronic superexchange models describing the orbital order of 
$e_g$ electrons were suggested early on by Kugel and Khomskii,
\cite{Kug82} and are of great interest recently.\cite{Tok00} In 
cuprates, like in KCuF$_3$, they lead to enhanced quantum effects,
\cite{Fei97} while in manganites these models, which originate from 
several charge excitations due to either $e_g$ or $t_{2g}$ electrons, 
are richer and include interactions between large and more classical 
spins.\cite{Shi97,Mae98,Fei99,Oka02} These interactions, together with 
the Jahn-Teller (JT) effect, stabilize $A$-type antiferromagnetic (AF) 
phase coexisting with orbital order in undoped LaMnO$_3$. While the 
shape of the occupied orbitals in the low-temperature phase of 
LaMnO$_3$ is still controversial, the two-sublattice orbital alternation 
in an orbital ordered state within the FM planes of LaMnO$_3$ is well 
established.\cite{Rod98} When the spin dynamics in the FM planes of 
LaMnO$_3$ is frozen at low temperature, one can constrain an effective 
model to {\it purely orbital superexchange}\cite{vdB99} instead of 
considering more complex complete spin-orbital model.\cite{Fei99} The 
superexchange orbital interactions favor then alternating directional 
($3z^2-r^2$-like) and planar ($x^2-y^2$-like) orbitals along every 
cubic direction.\cite{vdB99}

A considerable simplification is allowed in the FM phase at finite 
doping and at $T=0$, where the spins are aligned and an effective 
$t$-$J$-like charge-orbital model is sufficient. Experimentally, at 
doping higher than $x\simeq 0.10$ one finds a FM insulating phase, 
followed by a metallic phase.\cite{Bio01} The microscopic reasons of 
this behavior are intriguing --- it contradicts the usual spin 
polaronic picture of a FM phase. A puzzling competition between the 
insulating and metallic behavior within the FM phase was also reported 
for La$_{0.88}$Sr$_{0.12}$MnO$_3$.\cite{End99}  

Here we will concentrate on the generic behavior due to $e_g$ {\it 
orbital degrees of freedom} in a FM phase, and we will treat the spin 
dynamics classically, as usually done in the double exchange model.
\cite{Zen51,Bri99} This approach is complementary to focusing on the 
quantum effects in double exchange, which was presented recently.
\cite{Fes01} Our aim is to study the correlation functions and spectral 
properties of a one-dimensional (1D) orbital $t$-$J$ model by exact 
diagonalization at zero temperature ($T=0$), and by classical Monte 
Carlo simulations of spin correlations at finite temperature in order 
to establish the consequences of orbital order and orbital dynamics in 
doped manganites. In the strongly correlated regime the charge dynamics 
couples to orbital excitations and can be described in terms of the 
orbital $t$-$J$ model for doped manganites.\cite{Bri00,Bal02,Ole02} 
As in the spin $t$-$J$ model, a doped hole moves in an orbital ordered 
state by dressing itself with orbital excitations.\cite{Bri00} However, 
the structure of the quasiparticle (orbital polaron\cite{Bal02}) is here 
quite different from that derived from the $t$-$J$ model,\cite{Ram93} 
as the orbital model is more classical. 

As a reference system we use a 1D spin $t$-$J$ model for which the 
behavior of a single hole is well understood. A hole created in a 1D 
N\'eel state is mobile and may be thought as decaying into a magnetic 
domain wall. In fact, it gives a charged domain wall already after a 
single hop and leaves behind a solitonic defect from which it separates 
and next propagates independently. This is the simplest visualization 
of hole-spin separation in a 1D system. The familar string picture and 
quasiparticles on the energy scale of $\sim 2J$ are then recovered when 
a staggered magnetic field is applied and suppresses domains of 
reversed spins.\cite{Bon92,Wro96} 

We show below that an analogous phenomenon of hole-orbital separation
does not occur for the orbital degrees of freedom, but instead a single 
hole is trapped in a 1D chain when orbitals alternate. This explains why 
an insulating behavior may extend to finite doping. A staggered field 
has a well established physical origin in this case, and could follow 
from the frozen JT modes with alternating oxygen distortions along the 
chain itself, and on the bonds perpendicular to its direction, when the 
chain is embedded within a three-dimensional (3D) crystal. This 
interaction may play also an important role for the transport properties 
of lightly doped manganites and is difficult to separate from on-site 
Coulomb interaction $U$.\cite{Ben99} Therefore, it was even suggested 
that models including Hund's exchange $J_H$ between conductions 
electrons and $t_{2g}$ spins, but neglecting Coulomb interaction $U$, 
could capture the essential physics of manganites.\cite{Hot00} We will 
investigate below the JT term which leads to a staggered field acting on 
the orbitals. Note, however, that the 1D case is special as static JT 
distortions, preventing fluctuations of occupied orbitals, suppress 
completely hole motion along the chain, while in a two-dimensional (2D) 
model they lead instead to an enhanced coherent component in the hole 
motion mainly due to suppression of incoherent processes.\cite{Bal02} 

Another feature characteristic of the orbital physics and different 
from spin models is that the lattice responds to the doping. As pointed 
out by Kilian and Khaliullin,\cite{Kil99} the breathing motion of the 
MnO$_6$ octahedra provides a strong tendency towards hole localization 
in the presence of singly occupied (almost) degenerate $e_g$ levels,
as shown in a 2D model.\cite{Bal02} A static hole polarizes then the 
orbitals occupied by $e_g$ electrons in its neighborhood, and this 
polarization is expected to happen in addition to the effects promoted 
by the orbital superexchange interactions. We also consider this 
interaction in the present 1D model and show that it simply renormalizes 
the superexchange interaction, and thus it could lead to qualitatively 
new features and be of more importance only in higher dimension.  

The paper is organized as follows. In Sec. II we present the effective 
$t$-$J$ orbital model for $e_g$ electrons moving along a 1D chain in a
FM plane of, e.g., La$_{1-x}$Sr$_x$MnO$_3$. In Sec. III this model is 
analyzed first qualitatively by comparing the energies of two phases: 
(i)insulating phase with localized $e_g$ polarons, and (ii) metallic 
phase with itinerant carriers. Next we analyze the correlation functions 
and the spectral properties of a finite chain of $N=14$ sites filled by 
up to five holes, and demonstrate a crossover from the insulating to 
metallic phase under increasing doping. At finite temperature $T$ we 
derive an effective $t$-$J$ orbital model, with electron hopping and 
orbital interactions depending on actual configuration of core spins
(Sec. IV). The interrelation between spin and orbital order, the 
correlations around a doped hole, and the evolution of spectral 
properties under increasing doping are next investigated using a 
combination of exact diagonalization with Monte Carlo simulations. 
Finally, we summarize the results and present general conclusions in 
Sec. V.

\section{Orbital $t$-$J$ Model}
\label{sec:otj}

We consider the 1D orbital $t$-$J$ model,
\begin{equation}
{\cal H}_0=H_t+H_J+H_{\rm JT},
\label{HtJ}
\end{equation}
obtained for $e_g$ electrons in FM manganites at $T=0$. The Hamiltonian 
(\ref{HtJ}) acts in the restricted Hilbert space without double 
occupancies. Due to the absence of the SU(2) symmetry, the kinetic 
energy in the $e_g$ band takes a form which depends on the used orbital 
basis. For the present 1D model it is most convenient to consider 
a chain along $c$ axis and to use the usual orbital basis 
$\{x^2-y^2,3z^2-r^2\}$, for which we introduce a compact notation,
\begin{equation}
\label{realorbs}
\textstyle{
|x\rangle\equiv \frac{1}{\sqrt{2}}|x^2-y^2\rangle, 
\hspace{0.7cm}
|z\rangle\equiv \frac{1}{\sqrt{6}}|3z^2-r^2\rangle.}
\end{equation}
A chain along $a$ or $b$ axis could also be analyzed using this basis,
\cite{Hot00} but in each case one obtains a simpler and more transparent 
interpretation of the results with a basis consisting of a directional 
orbital along this particular cubic axis, and an orthogonal to it planar 
orbital, e.g. $3x^2-r^2$ and $y^2-z^2$ for $a$ axis. 

The model given by Eq. (\ref{HtJ}) stands for a chain composed of 
Mn$^{3+}$ and Mn$^{4+}$ ions coupled by the effective hopping $t$, which 
originates from the transitions over oxygens. Due to the symmetry, the 
$|x\rangle$ electrons cannot move, and the hopping term $H_t$ allows 
only for intersite transitions between a site occupied by a $|z\rangle$ 
electron and an empty neighboring site,  
\begin{equation}
H_t=-t\sum_{i}
    ({\tilde c}_{iz}^{\dagger}{\tilde c}_{i+1,z}^{}
    +{\tilde c}_{i+1,z}^{\dagger}{\tilde c}_{iz}^{}),
\label{Ht}
\end{equation}
where an operator ${\tilde c}_{iz}^{\dagger}=c_{iz}^{\dagger}(1-n_{ix})$
creates a $|z\rangle$ electron when site $i$ is unoccupied by an 
$|x\rangle$ electron. The hopping $H_t$ describes thus spinless fermions 
with an orbital flavor in a restricted Hilbert space, in analogy to the 
original $t$-$J$ model in spin space,\cite{Cha77} but only one ($z$) 
component is here {\it itinerant\/}, while the other ($x$) one is {\it
immobile\/} and hinders the motion of $|z\rangle$ electrons along the 
chain. 

The superexchange in undoped LaMnO$_3$ is given by a superposition of
several terms which originate from charge excitations due to either 
$e_g$ or $t_{2g}$ electron hopping\cite{Fei99} --- they are discussed 
in Sec. \ref{sec:efftJ}. While the $t_{2g}$ part is AF, the $e_g$ terms 
favor either FM or AF spin order on a bond $\langle ij\rangle$, 
depending on the pair of occupied $e_g$ orbitals at sites $i$ and $j$. 
For the realistic parameters of LaMnO$_3$, one finds the $A$-type AF 
order in the ground state, with FM planes staggered along the third 
direction. Taking a cubic $c$ direction within a FM plane, the 
superexchange expression simplifies enormously.\cite{vdB99} Treating 
large $S=2$ spins at Mn$^{3+}$ ions classically, all AF terms drop out 
at $T=0$, and the remaining {\it orbital superexchange} favors 
alternating $e_g$ orbitals.\cite{noteaf} Therefore, the superexchange 
interactions reduce then to the purely orbital interactions which favor 
alternating directional ($3z^2-r^2$-like) and planar ($x^2-y^2$-like) 
orbitals along every cubic direction.\cite{vdB99} In the present 1D 
model one finds 
\begin{equation}
H_J=2J\sum_{i} \Big( T_{i}^zT_{i+1}^z
   -\textstyle{\frac{1}{4}}{\tilde n}_{i}{\tilde n}_{i+1}\Big),
\label{HJ}
\end{equation}
where operators
\begin{equation}
\textstyle{
T_{i}^z=\frac{1}{2}\sigma_{i}^z=\frac{1}{2}(n_{ix}-n_{iz})},
\label{Tz}
\end{equation}
stand for orbital pseudospins $T=1/2$, with two eigenstates defined in 
Eq. (\ref{realorbs}), and ${\tilde n}_{i}=
{\tilde c}_{ix}^{\dagger}{\tilde c}_{ix}^{}+
{\tilde c}_{iz}^{\dagger}{\tilde c}_{iz}^{}$ is an electron number 
operator in the restricted Hilbert space. The superexchange constant 
$J=t^2/\varepsilon(^6\!A_1)$ is then given by the high-spin excitation 
energy $\varepsilon(^6\!A_1)$.\cite{Fei99}  

The last term in Eq. (\ref{HtJ}) stands for the staggered field induced 
by the cooperative JT effect, 
\begin{equation}
H_{\rm JT}=2E_{\rm JT}\sum_{i}\exp(i\pi R_i) T_{i}^z,
\label{HJT}
\end{equation}
considered also in Ref. \onlinecite{Hot03}, and
supporting the alternating orbital order. It gives an energy gain 
$E_{\rm JT}$ per site in the ground state of an undoped 1D chain, and 
follows from the alternating oxygen distortions around manganese ions 
in LaMnO$_3$.\cite{Mil96} The present simplified form [Eq. (\ref{HJT})] 
of a general expression \cite{Bal00}, which depends on the type of 
oxygen distortion, is sufficient in the 1D model. 

At finite doping $x=1-n$, where $n$ is an average $e_g$ electron number 
per site, the present $t$-$J$ model gives an interesting problem, with 
competing tendencies towards orbital alternation in insulating state on
one hand, described by orbital correlations,
\begin{equation}
T(n)=\langle T_i^zT_{i+n}^z\rangle.
\label{oo}
\end{equation}
and uniform $|z\rangle$ polarization in metallic state on the other.
The nature of the ground state obtained at finite doping is best 
investigated by considering a few characteristic correlation functions, 
containing information about the orbital state and about the orbital 
order between two nearest neighbor sites, both at distance $n$ from a 
hole, and about the orbital correlations across a hole. To optimize
them we introduce:   
\begin{eqnarray}
\label{ho}
P(n)&=&\langle {\bar n}_iT_{i+n}^z\rangle,                 \\
\label{hoo}
R(n)&=&\langle {\bar n}_iT_{i+n}^zT_{i+n+1}^z\rangle,      \\
\label{oho}
Z&=&\langle T_{i-1}^z {\bar n}_iT_{i+1}^z\rangle,
\end{eqnarray}
where ${\bar n}_i$ is a hole number operator at site $i$,
\begin{equation}
\label{nhole}
{\bar n}_i=1-{\tilde c}_{iz}^{\dagger}{\tilde c}_{iz}^{}
            -{\tilde c}_{ix}^{\dagger}{\tilde c}_{ix}^{}.
\end{equation}
These correlation functions will be discussed in Secs. III and IV.
Note that the kinematical constraint gives $P(0)=R(0)=0$. 

We compare the orbital correlations and the spectral properties 
obtained for the orbital chain, with these found for a 1D $t$-$J$ spin 
model for $S=1/2$ spins,
\begin{eqnarray}
{\cal H}_{tJ}&=&-t\sum_{i\sigma}
    ({\tilde c}_{i\sigma}^{\dagger}{\tilde c}_{i+1,\sigma}^{}
    +{\tilde c}_{i+1,\sigma}^{\dagger}{\tilde c}_{i\sigma}^{})\nonumber \\
&+&2J\sum_{i}\Big({\vec S}_i\cdot {\vec S}_{i+1}
   -\textstyle{\frac{1}{4}}{\tilde n}_{i}{\tilde n}_{i+1}\Big)\nonumber \\
&+&2h_s\sum_{i}\exp(i\pi R_i)S_i^z,
\label{spintJ}
\end{eqnarray}
where 
${\tilde c}_{i\sigma}^{\dagger}=c_{i\sigma}^{\dagger}(1-n_{i,-\sigma})$ 
is a creation operator of an electron with spin $\sigma$ at site $i$ in 
the restricted Hilbert space. In contrast to the orbital model, here 
the electrons with {\it both spin flavors are mobile\/} and exchange 
between the sites $i$ and $i+1$, so the interaction is a scalar product 
${\vec S}_i\cdot {\vec S}_{i+1}$ instead of the Ising term 
$T_i^zT_{i+1}^z$. For convenience, we use the same units as in Eqs. 
(\ref{HJ}) and (\ref{HJT}); here $J=2t^2/U$ with $U$ being the 
excitation energy. The staggered field $\propto h_s$ simulates the 
long-range spin order present in a 2D model,\cite{Bon92,Wro96} and 
plays a similar role to the JT field in the present orbital model. 
For doped spin chain we consider analogous correlation functions to: 
orbital-orbital (\ref{oo}), hole-orbital (\ref{ho}), orbital order at 
distance $n$ from the hole (\ref{hoo}), and around a hole (\ref{oho}), 
with spin operators $S_i^z$ in place of $T_i^z$ [Eq. (\ref{Tz})]. 

In the next Section we report the results obtained by Lanczos 
diagonalization of an $N=14$ orbital chain with periodic boundary 
conditions and different electron filling at $T=0$, and compare them to 
more familiar spin physics. We have verified that the results concerning 
the crossover to metallic phase are representative and do not depend on 
the chain length in any significant way.

\section{Numerical results at $T=0$}
\label{sec:num}

\subsection{Analytic estimation of the crossover from orbital polarons
            to itinerant electrons}
\label{sec:phd}

The ground state of the orbital $t$-$J$ model (\ref{HtJ}) depends on 
two parameters: $J/t$ and $E_{\rm JT}/t$, and on hole doping $x=1-n$. We 
first investigate the ground state (at $T=0$) changing these parameters. 
A realistic value of $t\sim 0.4-0.5$ eV was estimated for manganites 
using the charge-transfer model.\cite{Fei99} Taking this as an energy
unit, and the spectroscopic value for the energy of the high-spin
excitation $\varepsilon(^6A_1)=U-3J_H\simeq 3.8$ eV, this leads to 
$J/t\simeq 1/8$. We will consider also $E_{\rm JT}>0$, which promotes
localized behavior of $e_g$ electrons. 

\begin{figure}
\includegraphics[width=7.2cm]{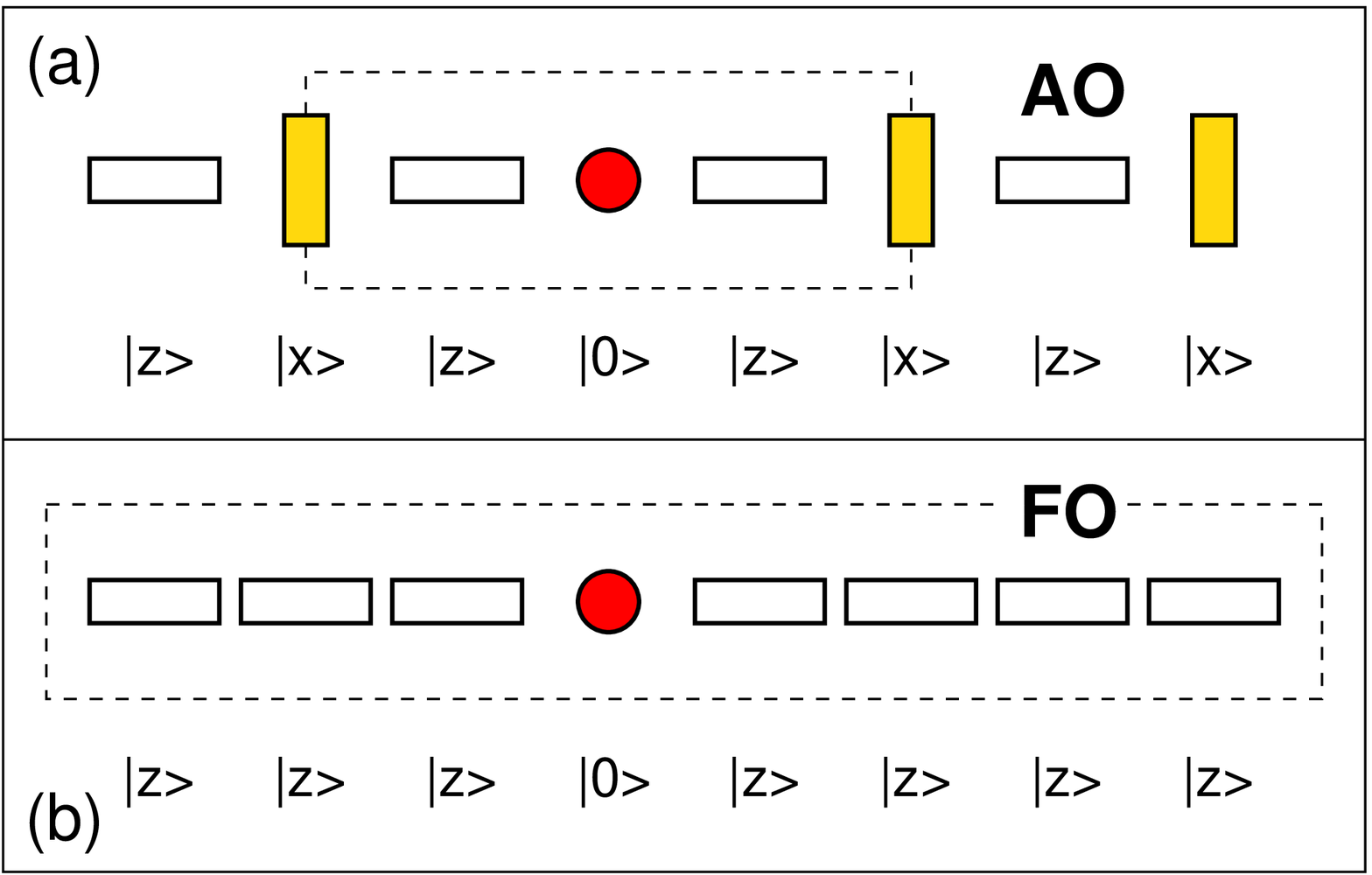}
\vskip .2cm
\includegraphics[width=7.7cm]{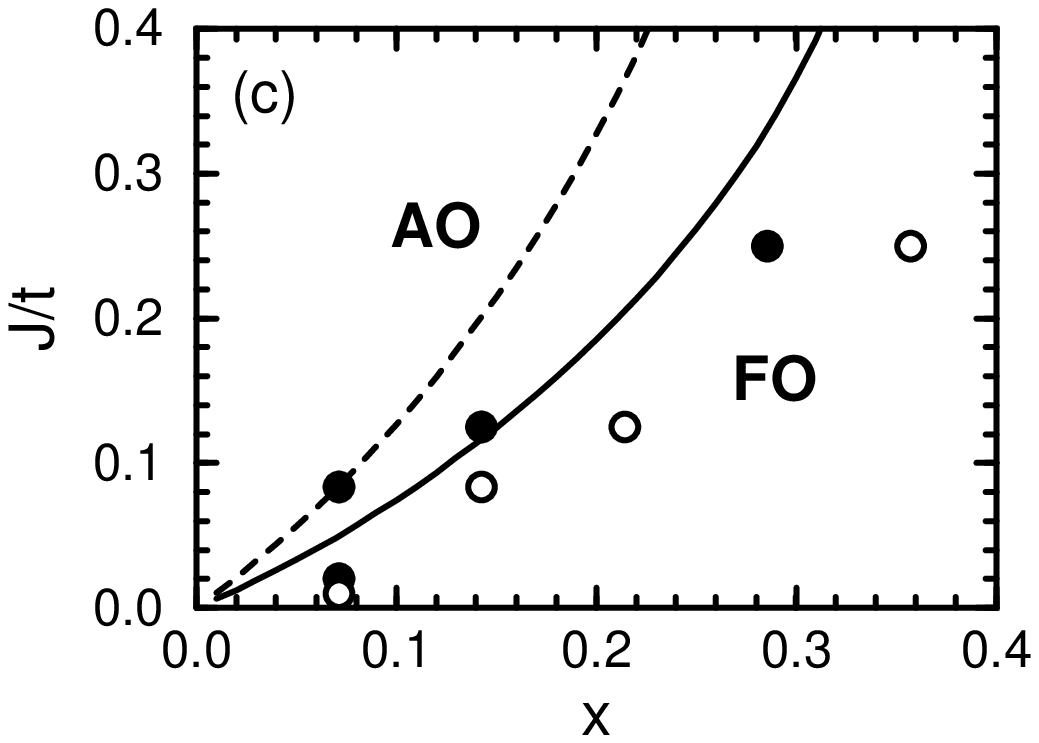}
\caption{(color online)
Competition between localized holes within an alternating orbital (AO) 
phase, with staggered localized $|x\rangle$ (shadded vertical boxes) 
and itinerant $|z\rangle$ (empty horizontal boxes) orbitals (a), and 
itinerant holes in ferro orbital (FO) $|z\rangle$ phase (b), in a 1D 
chain along $c$ axis. A hole (shadded circle) is confined to a cluster 
of three sites in the AO state, while it is delocalized over the entire 
chain in the FO state, as indicated by dashed boxes. Part (c) shows a 
qualitative phase diagram of the 1D orbital $t$-$J$ model (\ref{HJT}), 
with a critical concentration $x_c$ separating the AO and FO phases, 
as obtained for $E_{\rm JT}=0$ and: 
$\Delta=0$ (solid line), 
$\Delta=t$ (dashed line). 
Filled (empty) circles in (c) show the AO (FO) states found by exact
diagonalization of an $N=14$ chain.
}
\label{fig:phd}
\end{figure}

At half-filling one finds alternating $|x\rangle/|z\rangle$ orbitals 
in the ground state at $J>0$, with classical intersite correlations
$T(n)=(-1)^n/4$. If a single hole is then doped to orbitally ordered 
state at an $|x\rangle$ site, it can delocalize within a box consisting 
of three sites, as each neighboring $|z\rangle$ electron can 
interchange with the hole [see Fig. \ref{fig:phd}(a)]. This state is 
therefore favored over doping at $|z\rangle$ site, and one can easily 
determine the energies of a doped hole in an antibonding and bonding 
state,    
\begin{equation}
\label{eh}
E_{1h,\pm}=\frac{1}{2}{\bar J}
           \pm\frac{1}{2}\Big({\bar J}^2+8t^2\Big)^{1/2}, 
\end{equation}
with ${\bar J}=J+2E_{\rm JT}$ standing for an excitation energy of the
configuration with a hole moved to a left (right) site within a 
three-site cluster shown in Fig. \ref{fig:phd}(a). 

As long as the holes may be doped into separated three-site units 
(for $x\leq 0.25$), the total energy (per site) of an insulating phase 
follows from the weighted contributions of the undoped orbital ordered 
regions, and doped holes occupying the bonding states of individual
clusters, with energy $E_{1h,-}$ (\ref{eh}), 
\begin{equation}
\label{ei}
E_I=-(1-x)E_{\rm JT}-(1-2x)J+xE_{1h,-}. 
\end{equation}
The energy of a metallic phase, which contains only itinerant 
$|z\rangle$ electrons for $n>0.5$ [see Fig. \ref{fig:phd}(b)], 
\begin{equation}
\label{em}
E_M=\frac{1}{\pi}\int_{k<\pi/2}\varepsilon_{k,-}
   +\frac{1}{\pi}\int_{k<k_F}  \varepsilon_{k,+}, 
\end{equation}
is obtained by integrating over the occupied one-particle band states
$\varepsilon_{k,\pm}$. At finite $E_{\rm JT}$ one finds,  
\begin{equation}
\label{ek}
\varepsilon_{k,\pm}=\pm\sqrt{E_{\rm JT}^2+4t^2\cos^2 k}. 
\end{equation}
The superexchange $\propto J$ does not contribute to $E_M$. 

It is now of interest to compare the energy of two extreme situations: 
(i) an insulating phase with holes localized within three-site clusters,
[Eq. (\ref{ei})], with (ii) a metallic (itinerant) phase [Eq. 
(\ref{em})]. One finds that at fixed $J$ and $E_{\rm JT}=0$, an 
{\it insulator-metal transition} takes place when the hole concentration 
$x$ increases. A critical concentration for this transition increases 
with increasing $J/t$ [Fig. \ref{fig:phd}(c)]. The data points obtained 
from the exact diagonalization of an $N=14$ chain agree with the 
analytic estimation at $J/t=1/12$ and $1/8$, while for a larger value of 
$J/t=1/4$ the region of the insulating AO phase is more extended in the 
chain. Note, however, that the present analytic estimate is anyway only 
qualitative as the energy of the AO phase can be evaluated using a 
superposition of holes confined to three-site clusters only up to 
$x=0.25$. An extended region of stability of the AO phase results here 
from larger clusters of the itinerant phase which are still separated by 
immobile $|x\rangle$ states; such larger itinerant units occur already 
for $x<0.25$ at low values of $J/t$, as the gain in the kinetic energy 
for a hole moving within a larger cluster approaches fast the metallic 
limit with increasing cluster size, and may be easily compensated when 
a few exchange bonds are created. Indeed, the data show [see Fig. 
\ref{fig:phd}(c)] that the ground state for $x=1/14$ remains insulating 
with two $|x\rangle$ electrons at $J=0.02t$, and becomes metallic only 
for $J<0.02t$. 

A simple estimation of the metal-insulator transition is also possible
when a polaronic polarization around a hole\cite{Kil99} is included 
in Eq. (\ref{HtJ}), 
\begin{equation}
\label{HDelta}
H_\Delta=-\Delta\sum_i {\bar n}_i
          ({\tilde n}_{i-1,z}+{\tilde n}_{i+1,z}), 
\end{equation}
where ${\tilde n}_{jz}={\tilde c}_{i\pm 1,z}^{\dagger}
{\tilde c}_{i\pm 1,z}^{}$ is the electron number operator in the 
restricted space without double occupancies, and ${\bar n}_i$ is a hole 
number operator (\ref{nhole}). At $E_{\rm JT}=0$ a similar expression to 
Eq. (\ref{eh}) is then obtained, with ${\bar J}=J+\Delta$, and the range
of an insulating AO phase shrinks in the phase diagram of Fig. 
\ref{fig:phd}(c). In fact, the polarization around doped holes is 
optimized in a metallic phase, and this result is special to the 1D 
model. Therefore, we shall not consider this interaction further, as 
it gives qualitatively the same results as the orbital $t$-$J$ model 
at $\Delta=0$, but with a somewhat reduced value of $J$. 

At finite JT energy $E_{\rm JT}$ the situation changes in a drastic way.
Even for relatively small $E_{\rm JT}\simeq 0.25t$ the AO phase is 
stabilized in the entire range of doping shown in Fig. \ref{fig:phd}(c). 
Note that even somewhat higher values could be more appropriate for 
realistic 3D manganites.\cite{Kil99,Bal00} This suggests that the JT 
effect may indeed play a very important role in manganites and stabilize 
an insulating phase with orbital order in a broad regime of doping; we 
address this question in Sec. \ref{sec:jt}.

\subsection{Orbital order at half filling}
\label{sec:undoped}

\begin{figure}
\includegraphics[width=7.7cm]{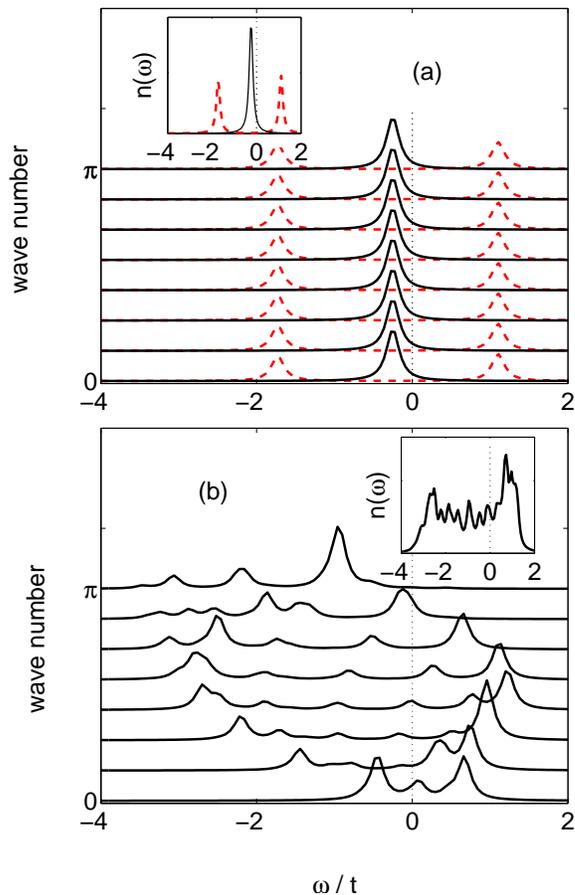}
\caption{(color online)
Spectral functions $A(k,\omega)$ at half filling for an $N=14$ site 
chain: 
(a) orbital model, with hole excitations within occupied 
$|z\rangle/|x\rangle$ orbital shown by solid/dashed lines, 
respectively; 
(b) spin $t$-$J$ model. Insets show the density of states $n(\omega)$.
Parameters: $J=0.125t$, $E_{\rm JT}=0$, $h_s=0$. We assumed a finite 
broadening of the peaks by $\delta=0.1t$.}
\label{fig:n=1}
\end{figure}

The superexchange interaction in the orbital model (\ref{HJ}) is 
Ising-type, and therefore the orbital order in the undoped system at
finite $J$ is perfect, with alternating occupied $|x\rangle$ and 
$|z\rangle$ orbitals along the chain, and 
$\langle T_{i}^zT_{i+1}^z\rangle=-0.25$. The classical character of 
this ground state is reflected in the one-hole excitation spectra. If a 
single hole is added at half filling ($n=1$), it may be doped either at 
an $|x\rangle$ or at a $|z\rangle$ site. A hole doped at a $|z\rangle$ 
site is immobile and the energy of the final state is higher by $2J$ 
than that of the initial state, as two exchange bonds are removed. This 
excitation appears as a localized peak at hole binding energy $-2J$ in 
Fig. \ref{fig:n=1}(a). In contrast, when a hole replaces an $|x\rangle$
electron at site $i$, the resulting state is not an eigenstate of 
${\cal H}_0$, the hole delocalizes over a three-site cluster including 
also occupied $|z\rangle$ orbitals at neighboring sites $i-1$ and $i+1$ 
[Fig. \ref{fig:phd}(a)], and the bonding and antibonding hole states 
(\ref{eh}) contribute. They give two maxima in the excitation spectrum 
of Fig. \ref{fig:n=1}(a), with energies $E_{1h,\pm}-2J$. As the hole 
cannot hop over the sites occupied by $|x\rangle$ electrons, these two 
states are again localized within a cluster, and lead to 
$k$-independent maxima in the spectral function. 

Strictly speaking, three-site terms, similar to those known from the 
derivation of the spin $t$-$J$ model,\cite{Cha77} occur as well in the
orbital model. One might expect that such terms, $\propto J
{\tilde c}_{i\pm 1,z}^{\dagger}{\tilde n}_{ix}{\tilde c}_{i\mp 1,z}$, 
would change the spectral functions as a hole created at $|z\rangle$ 
site could then hop to its second neighbors and interchange with 
immobile electrons in $|x\rangle$ orbitals. However, we have verified 
that these processes are of importance only for $J\sim t$, but do not 
lead to any significant changes of the spectral properties in the 
physically interesting regime of $J\lesssim 0.25t$. Therefore, they are 
neglected in what follows.  

\begin{figure}
\includegraphics[width=7.7cm]{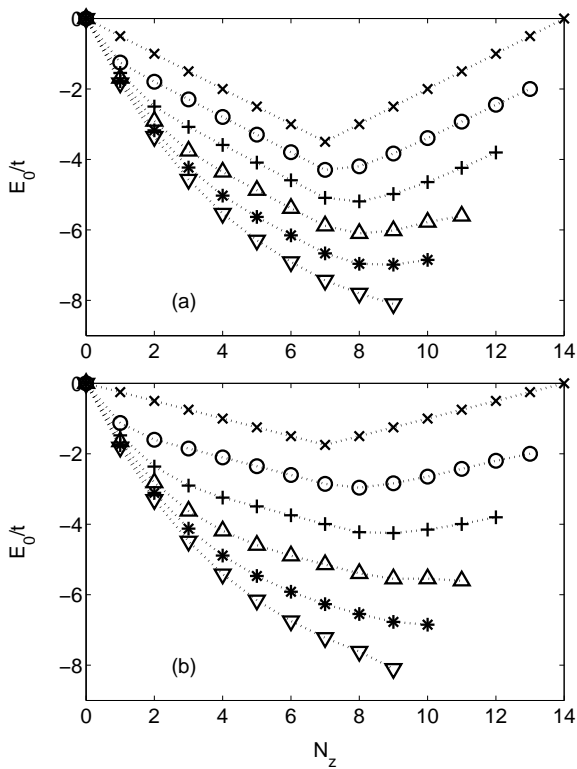}
\caption{
Ground state energy $E_0$ as a function of the number of itinerant 
electrons $N_z$ for 0 up to 5 holes (from top to bottom) added to a 
half-filled $N=14$ orbital chain, for $E_{\rm JT}=0$, and for: 
(a) $J=0.25t$, and   
(b) $J=0.125t$. 
}
\label{fig:enz}
\end{figure}

In contrast, the spectral functions obtained with the usual spin $t$-$J$
model show a dispersive feature at the highest energy, a similar weaker
dispersive feature at low energy, and incoherent spectral intensity 
between them [Fig. \ref{fig:n=1}(b)]. The two dispersive features have 
a periodicity of $\pi$ instead of $2\pi$ obtained for free electrons, 
are broadened and rather incoherent. Unlike in the 2D case where a hole 
is trapped and can move only by its coupling to quantum fluctuations in 
the spin background, leading to a quasiparticle behavior with a 
dispersion on the energy scale $\propto J$,\cite{Mar91} 
the string potential is absent in a 1D chain. The 1D model shows
spin-charge separation, because a hole may propagate after creating and 
leaving behind a single solitonic defect with two spins of the same 
direction next to each other.\cite{Zac98,Sen00,Aic03} However, the
chain used in our case is too small for the spin-charge separation to be
clearly visible. A more coherent component can be induced by introducing 
the coupling between the moving charge and spin fluctuations, which 
occurs in the presence of an external staggered field generating the 
string potential.\cite{Bon92,Wro96} We address this issue in Sec.
\ref{sec:jt}.

The present $N=14$ spin chain gives at half filling 
$\langle S_i^zS_{i+1}^z\rangle=-0.1491$, a value being already quite 
close to the exact result $-0.1477$ for an infinite Heisenberg chain. 
Therefore, we expect that the results presented for the more classical 
orbital chain with partly localized wavefunctions are at least of the 
same quality, and suffer even less from finite size effects.

\subsection{Correlation functions at finite doping}
\label{sec:co}

First we consider the total energy $E_0=\langle{\cal H}_0\rangle$ as a 
function of the number of $|z\rangle$ electrons $N_z$ at each doping 
$x$, and determine the actual distribution of $e_g$ electrons  
in the ground state. The density of $e_g$ electrons in two orbitals, 
$n_x$ and $n_z$, depends on the ratio $J/t$. At half filling ($x=0$) 
one finds $n_x=n_z=0.5$, and the occupied orbitals alternate. At large 
$J/t\sim 1$ one expects that the orbital order is still close to perfect 
in the regions which separate orbital polarons, with the densities
$n_z\simeq 0.5$ and $n_x\simeq 0.5-x$. On the contrary, in the limit of 
$J/t\to 0$ a single hole (in a finite chain) suffices to destabilize the 
orbital order, causing a transition to an itinerant (FO) state with 
$n_z=1-x$, and $n_x=0$.

\begin{figure}
\includegraphics[width=7.7cm]{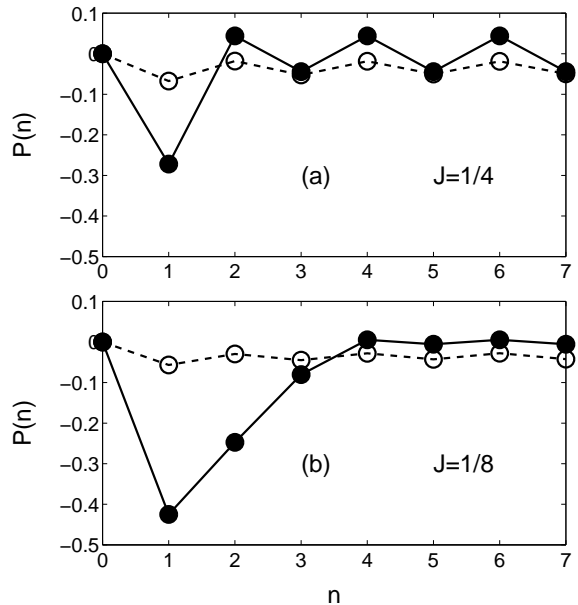}
\caption{
Polarization of the orbital (filled circles, full lines) and spin (open 
circles, dashed lines) background $P(n)$ (\ref{ho}) at distance $n$ from 
a single hole doped to a half-filled $N=14$ chain, as obtained for 
the ground state of the orbital/spin $t$-$J$ model without staggered 
fields ($E_{\rm JT}=0$, $h_s=0$), and for: 
(a) $J=0.25t$ ($N_z=7$), and  
(b) $J=0.125t$ ($N_z=8$).  
}
\label{fig:pn}
\end{figure}

The energy obtained for an $N=14$ chain filled by $9\leq N\leq 14$ 
electrons for two values of $J=0.25t$ and $0.125t$ are shown in Fig. 
\ref{fig:enz}. The tendency towards electron delocalization is quite 
distinct already for a higher value of $J=0.25t$, with the minimum of 
$E_0$ moving to $N_z>7$ with increasing doping $x$. The number of 
$|z\rangle$ electrons {\it increases\/} by one (to $N_z=8$) at the 
electron filling of $N_e=12$ and 11 electrons. At $N_e=10$ there remains 
just a single $|x\rangle$ electron which still blocks the hopping along 
the chain, while at $N_e=9$ all electrons are in $|z\rangle$ orbitals, 
and one obtains a metallic state. This transition to a metallic state is 
faster at a lower value of $J=0.125t$ [see also Fig. \ref{fig:phd}(c)].
Here adding a hole increases simultaneously the number of $|z\rangle$ 
electrons by one, and one finds $N_z=8$ and $N_z=9$ for $N_e=13$ and 
$N_e=12$, respectively, while doping by three holes gives already a 
metallic state with $N_z=N_e=11$.

The correlation functions in the ground state obtained after doping the 
chain by a single hole are very transparent at a higher value of 
$J=0.25t$, and are easily accessible by looking at the correlation 
functions (\ref{ho})--(\ref{oho}). As shown schematically in Fig. 
\ref{fig:phd}(a), a hole replacing $|x\rangle$ electron is then confined 
to a three-site cluster, which can be deduced from the hole-orbital 
correlation function $P(n)$ [see Fig. \ref{fig:pn}(a)]. First of all, 
at the nearest-neighbor of a hole one finds preferably a $|z\rangle$ 
electron, with $P(1)<-0.25$ indicating that a hole spends more time 
at a central site than at either outer site of the three-site cluster 
[$P(1)=-0.25$ would correspond to the bonding state at $J=0$ and 
$P(1)=-0.5$ to a static hole at a central site]. This also causes weak 
alternation of $P(n)$ with increasing $n$ for further neighbors
($n\geq 2$). This result is different again from the spin $t$-$J$ model, 
where almost no preference for the spin direction is found already at 
the nearest neighbor of a doped hole. Here the oscillations between even 
and odd neighbors are also much weaker as a hole is now delocalized, and 
one is averaging over several different configurations when $P(n)$ is 
evaluated. They correspond to the domains with opposite spins, and weak 
oscillations result here only from holon-spinon correlations. 

\begin{figure}
\includegraphics[width=7.7cm]{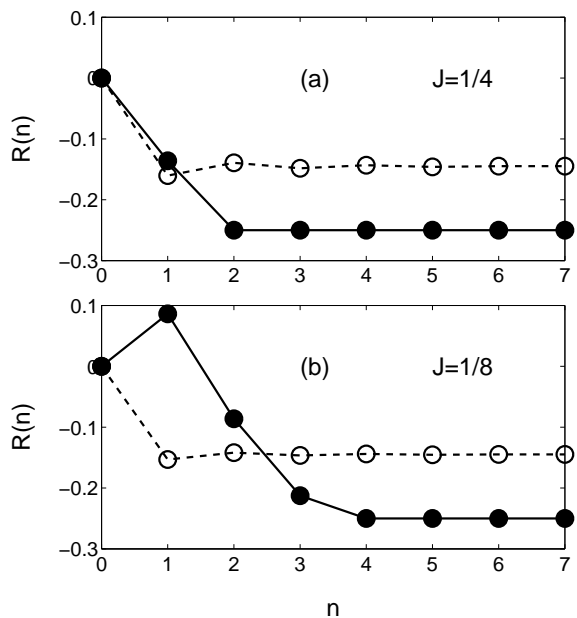}
\caption{
Orbital correlations $R(n)$ (\ref{hoo}) at distance $n$ from a single 
hole doped to a half-filled $N=14$ chain (filled circles, full lines), 
as obtained for the orbital $t$-$J$ model for $E_{\rm JT}=0$, and: 
(a) $J=0.25t$, and  
(b) $J=0.125t$. 
Spin correlations obtained in the 1D $t$-$J$ model ($h_Q=0$) is shown 
for comparison by open circles and dashed lines. 
}
\label{fig:rn}
\end{figure}

The ground state obtained for one hole at $J=0.125t$ is qualitatively 
different, as doping by one hole generates a single orbital flip and the 
number of itinerant $|z\rangle$ electrons increases to $N_z=8$. Fig. 
\ref{fig:pn}(b) shows that this defect in the otherwise perfect orbital 
order occurs close to the hole, increasing the island over which the 
hole can delocalize, extending now over five sites. 
Hence, one finds $P(1)\simeq -0.44$ and $P(2)\simeq -0.25$, and only 
starting from the fourth neighbor this correlation is weakly positive. 
In contrast, decreased value of $J$ does not cause any significant
change in the $P(n)$ correlations for the spin $t$-$J$ model --- they 
are only slightly weaker than for $J=0.25t$.

The orbital order, measured by second correlation function $R(n)$ 
(\ref{hoo}), remains perfect at a sufficient distance from the doped 
hole, with $R(n)=-0.25$ [Fig. \ref{fig:rn}(a)]. In case of a higher 
value of $J=0.25t$, the order is perfect starting from $n=2$, while for 
$J=0.125t$ with a larger island of the itinerant phase, it starts only 
at $n=4$. In addition, the orbital correlation $R(1)$ is here positive
[Fig. \ref{fig:rn}(b)], as two $|z\rangle$ orbitals occur frequently 
next to each other both at the first and at second neighbor of the hole 
which polarizes a larger five-site cluster. This behavior shows that 
phase separation occurs here in the regime of small $J/t$, and its 
mechanism which originates from orbital physics is completely different 
from that known from the spin $t$-$J$ model at large $J/t>1$. 

\begin{figure}
\includegraphics[width=7.7cm]{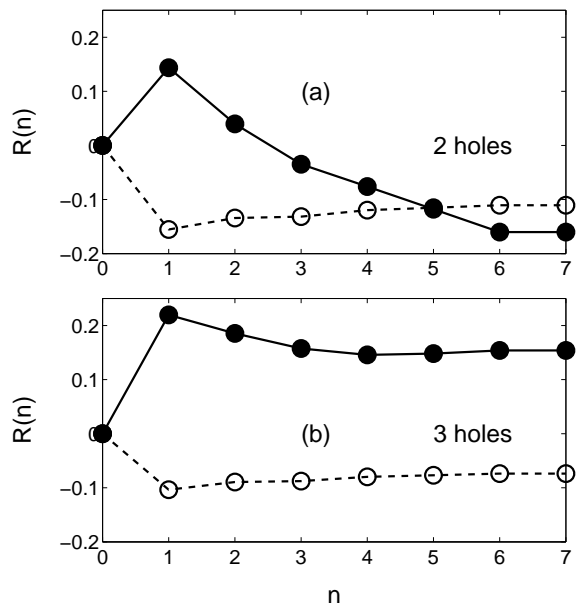}
\caption{
Orbital correlations $R(n)$ (\ref{hoo}) in a doped $N=14$ site chain 
(filled circles, full lines), as obtained for the orbital $t$-$J$ 
model at $J=0.125t$, $E_{\rm JT}=0$ for doping by: 
(a) two holes (insulating phase), and  
(b) three holes (metallic phase).
Spin correlations as in Fig. \ref{fig:rn}. 
}
\label{fig:rn3h}
\end{figure}

The spin $t$-$J$ model has practically the same spin-spin correlations
for both values of $J$. Due to quantum fluctuations in the Heisenberg 
1D chain, the correlation function $R(n)\simeq -0.14$ for $n>1$ [Fig. 
\ref{fig:rn}(a)] is much reduced from the classical value $R(n)=-0.25$ 
found for the orbital model. In fact, the long-range order is absent in 
the 1D spin chain, and the quantum fluctuations contribute to the energy 
(per site), which is much lower than in the orbital case, $e_0=J\langle 
{\vec S}_i\cdot {\vec S}_{i+1}\rangle=3J\langle S_i^zS_{i+1}^z\rangle$. 
The present $N=14$ spin chain gives at doping by one hole 
$\langle \bar{n}_iS_{i+n}^zS_{i+n+1}^z\rangle=-0.1448$ for $n=6,7$, a 
value which is only little reduced from $-0.1491$ found at half filling, 
showing that in spite of certain hole delocalization, spin intersite 
correlations are almost undisturbed by a single doped hole when the 
distance from it is sufficient.

\begin{figure}
\includegraphics[width=7.7cm]{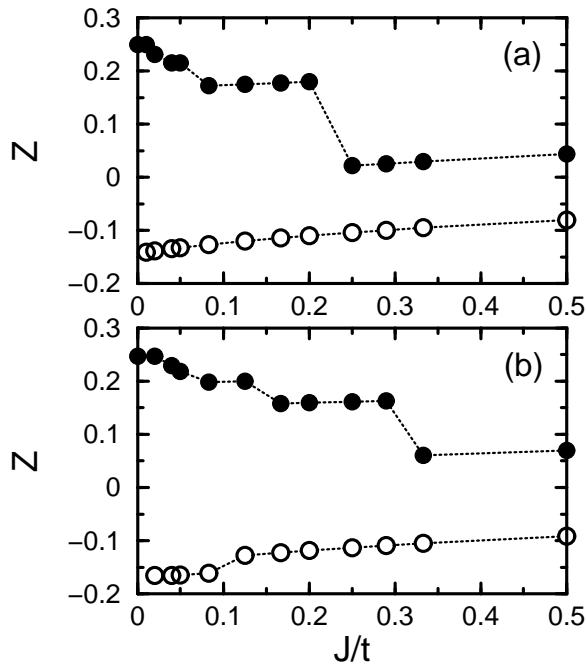}
\caption{
Orbital and spin order (filled and empty circles) around a hole $Z$ 
(\ref{oho}) as obtained for an $N=14$ chain with increasing $J/t$ 
doped by: 
(a) one hole, and (b) two holes. 
Other parameters as in Fig. \ref{fig:n=1}.
}
\label{fig:oho}
\end{figure}

When doping increases, a gradual crossover in $R(n)$ correlations 
towards a metallic chain with $|z\rangle$ orbitals occupied is found.
The region doped by holes extends for two holes over nine sites and
gives $R(n)<0$ starting from $n=3$ [Fig. \ref{fig:rn3h}(a)]. The 
correlation function $R(n)$ is averaged over several bonds, and the
values of $R(n)\simeq -0.17$ for $n=6,7$ result from a superposition of
the classical order ($\langle T_i^zT_{i+1}^z\rangle=-0.25$) in the 
insulating region and a positive value of 
$\langle T_i^zT_{i+1}^z\rangle$ correlations within the metallic 
cluster. When the insulating phase disappears at doping by three holes
[Fig. \ref{fig:rn3h}(b)], $R(n)>0$ shows that only $|z\rangle$ orbitals
are occupied. Three holes in the metallic phase avoid each other, and 
one finds a weak local minimum at $n=4$ in $R(n)$ correlations. The 
spin system gives instead negative $R(n)$, as here the hole hopping does 
not destroy the AF spin order. The correlation functions $R(n)$ for the 
spin $t$-$J$ model have similar values for either two or three holes, 
while local minima found at $n=7$ and $n=4$ indicate again 
characteristic distance between the holes in both cases.

The {\it polaronic\/} character of a doped hole is confirmed by the 
{\it positive\/} orbital correlation $Z>0$ between its neighbors 
(\ref{oho}). At $J\leq 0.01t$ the ground state is metallic and 
$Z=0.25$. Next, when $J/t$ increases, the hole is first trapped in a 
large metallic cluster ($Z\simeq 0.22$), with its size gradually 
decreasing down to a five-site cluster when $J\geq 0.08t$ --- then a 
hole is surrounded predominantly by two $|z\rangle$ electrons [Fig. 
\ref{fig:oho}(a)]. This ground state, with $N_z=8$ and $N_x=5$ 
electrons in the chain remains stable up to $J=0.20t$. At $J/t>0.20$ 
the ground state changes again to $N_z=7$ and $N_x=6$ which gives a 
small orbital polaron [see Fig. \ref{fig:phd}(a)]. Due to finite 
superexchange energy $J$, the probability that the hole is in the center 
of a three-site cluster, with two neighboring $|z\rangle$ electrons, 
is somewhat higher than that it occupies either of side atoms, with 
one $|z\rangle$ and one $|x\rangle$ occupied orbital next to it. As a 
result, the correlation function $Z\simeq 0.02$ is now weakly positive. 
For the filling by two holes one has again five different ground states
[Fig. \ref{fig:oho}(b)], with $Z=0.25$, $Z\simeq 0.22$, $Z\simeq 0.20$, 
$Z\simeq 0.16$, and $Z\simeq 0.07$ for increasing $J/t$. The transition
to two separated three-site polarons occurs here at a somewhat higher 
value of $J\sim 0.30t$ than the transition to a single three-site 
polaron for the doping by one hole. 

\begin{figure}
\includegraphics[width=7.7cm]{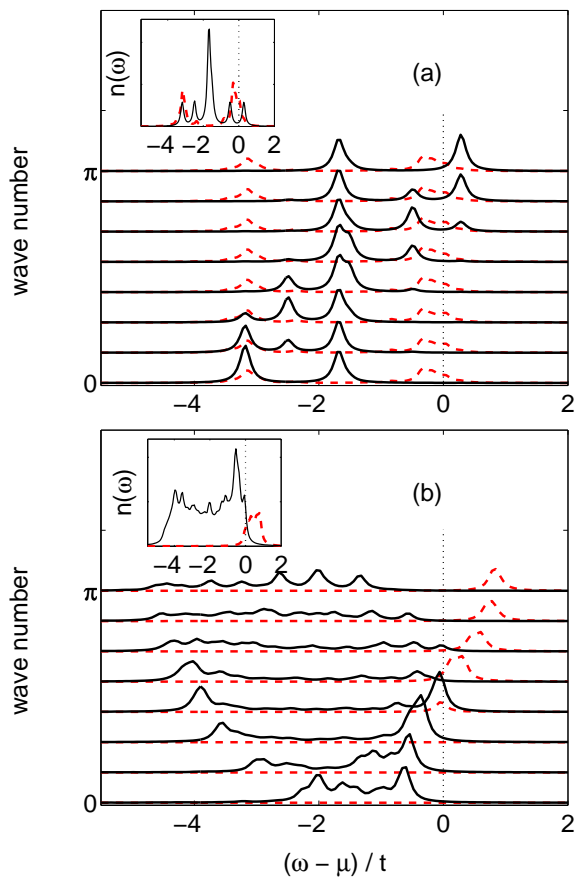}
\caption{(color online)
Spectral functions $A(k,\omega)$ for $N=14$ chains doped with one hole 
for $J/t=0.125$:
(a) orbital $t$-$J$ model (\ref{HtJ}) with $N_z=8$ and $N_x=5$ --- 
solid and dashed lines for $|z\rangle$ and $|x\rangle$ excitations;  
(b) spin $t$-$J$ model (\ref{spintJ}) with $N_{\uparrow}=6$ and 
$N_{\downarrow}=7$ --- solid and dashed lines for hole and electron 
excitations. Insets show the densities of states $n(\omega)$.
Other parameters and peak broadening as in Fig. \ref{fig:n=1}.
}
\label{fig:a1251h}
\end{figure}

In contrast, the spin correlations between two hole neighbors are {\it 
negative\/} in the spin $t$-$J$ model, both for one and for two holes 
(Fig. \ref{fig:oho}), showing the {\it solitonic\/} character of charge 
defect in the spin chain.\cite{Wro96} At low $J/t$ these correlations 
are more pronounced, while increasing $J/t$ leads to more delocalized 
holes and to weaker correlations $Z\simeq -0.10$ and $Z\simeq -0.12$ for 
the states doped either by one or by two holes at $J/t=0.1$. Somewhat 
stronger AF correlations in two-hole case confirm the solitonic 
character of charge defects in a 1D spin chain, with a soliton 
compensated by an antisoliton.

\subsection{Spectral functions at finite doping}
\label{sec:spectra}

Next we analyze the spectral functions obtained at finite doping.
Starting from the ground state of an orbital chain with a single hole, 
we found that in the region of very small $J$ ($J\sim 0.01t$ for $N=14$ 
chain) the spectral function of the orbital model shows a free 
propagation of a hole within the entire chain, and electrons fill 
$|z\rangle$ orbitals. This case (not shown) is however only of 
theoretical interest, and at $J\simeq 0.125t$, adequate for for 
LaMnO$_3$, one finds instead fairly localized spectra [Fig. 
\ref{fig:a1251h}(a)]. One recognizes the maxima which correspond to 
localized $|x\rangle$ excitations at $\omega-\mu\simeq -1.8t$, and the 
structures corresponding to the bonding and antibonding states of the 
$|x\rangle$ excitations at $\omega-\mu\simeq -3.3t$ and $-0.3t$. The 
spectra are $k$-dependent and the spectral weight moves to higher 
energies with increasing momentum $k$, following the tight-binding 
dispersion at $E_{\rm JT}=0$, $\varepsilon_{k-}=-2t\cos k$ (\ref{ek}). 
The total widths of the spectrum in Fig. \ref{fig:a1251h}(a) is close 
to $4t$, i.e., to full free-electron dispersion obtained for electrons 
in $|z\rangle$ orbitals.

\begin{figure}
\includegraphics[width=7.7cm]{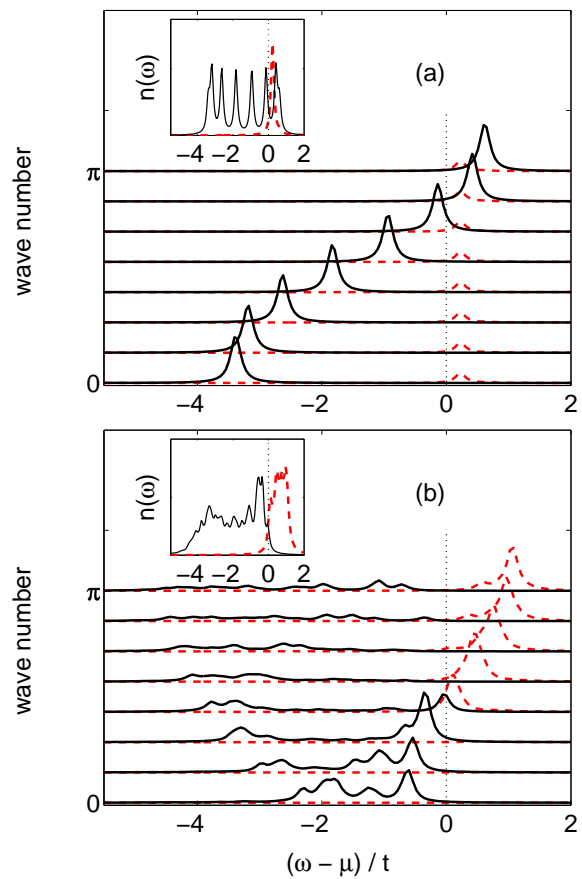}
\caption{(color online)
Spectral functions $A(k,\omega)$ for $N=14$ chains doped with
three holes for $J/t=0.125$:
(a) orbital $t$-$J$ model (\ref{HtJ}) with $N_z=11$ and $N_x=0$,  
(b) spin $t$-$J$ model (\ref{spintJ}) with $N_{\uparrow}=5$ and 
$N_{\downarrow}=6$.
The meaning of solid (dashed) lines, insets and other parameters are 
the same as in Fig. \protect\ref{fig:a1251h}.
}
\label{fig:a1253h}
\end{figure}

The spectrum obtained for spin $t$-$J$ model is drastically different.
A broadened quasiparticle band crossing the Fermi energy resembles a 
tight-binding dispersion with a reduced bandwidth [Fig. 
\ref{fig:a1251h}(b)]. For low momenta $k$, where a larger system would 
show spin-charge separation, the spectra are especially incoherent.
Most of the spectral intensity is in the incoherent part, but one
recovers a coherent propagation in the limit of $J\to 0$, which occurs 
in this case without polarizing the chain, as both spin flavors are 
mobile in this limit and the spin direction is irrelevant.

As we have shown in Fig. \ref{fig:enz}(b), doping $N=14$ chain by three 
holes is sufficient for the crossover to the metallic state for a 
realistic value of $J/t=1/8$. The spectral function $A(k,\omega)$ 
confirms the metallic behavior of the orbital chain for this filling, 
and the excitations in $|x\rangle$ orbitals contribute with a 
$k$-independent maximum only above the Fermi energy [Fig. 
\ref{fig:a1253h}(a)]. In contrast, the spectral functions in the spin 
$t$-$J$ model [Fig. \ref{fig:a1253h}(b)] are qualitatively very similar 
to those found at lower doping by one hole [Fig. \ref{fig:a1251h}(b)], 
but with more weight transferred now to the quasiparticle states above
the Fermi energy $\mu$ at $k>\pi/2$. The weight of the quasiparticle 
states for $\omega>\mu$ is enhanced, while the incohorent spectral 
weight is decreased for these values of $k$. 

\begin{figure}
\includegraphics[width=7.7cm]{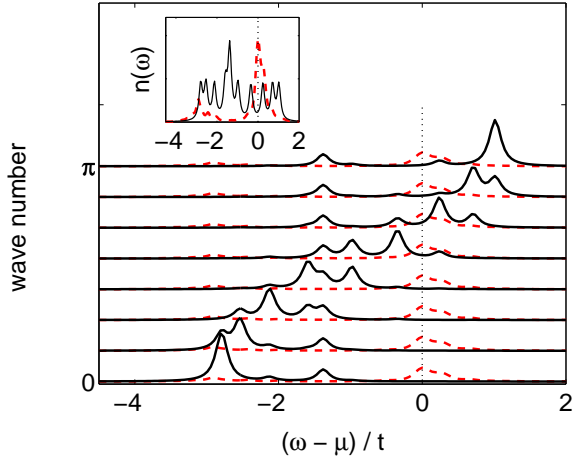}
\caption{(color online)
Spectral functions $A(k,\omega)$ for an $N=14$ orbital chain doped with
three holes for $J/t=0.25$. Solid and dashed lines for $|z\rangle$ and 
$|x\rangle$ excitations. Inset, other parameters and peak broadening as 
in Fig. \ref{fig:n=1}.
}
\label{fig:a2503h}
\end{figure}

The ground state of the orbital chain is insulating at a larger value 
of $J=0.25t$ [Fig. \ref{fig:enz}(a)]. Therefore, the spectral functions 
consist in this case of several incoherent features for each $k$ value,
with the first moment of the spectra following again the single-particle
dispersion (Fig. \ref{fig:a2503h}). The incoherent feature at 
$\omega\simeq\mu$, originating from excitations within $|x\rangle$ 
orbitals, is partly below the Fermi energy. This part of the spectral
weight follows from $|x\rangle$-hole excitations which are still 
possible at the filling by $N_z=9$ and $N_x=2$ electrons.

\subsection{Polarons induced by the JT effect}
\label{sec:jt}

\begin{figure}
\includegraphics[width=7.7cm]{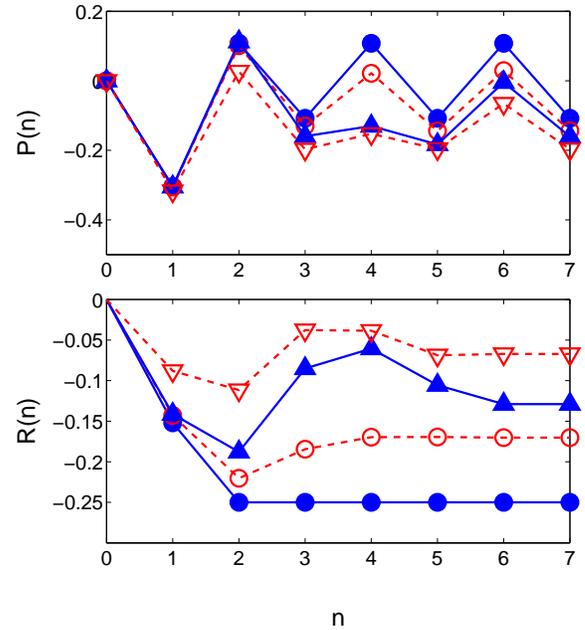}
\caption{
Evolution of orbital correlations with increasing hole doping, as 
obtained for $J=0.125t$ and $E_{\rm JT}=0.25t$:
(a) hole-orbital correlations $P(n)$, and
(b) hole-orbital-orbital correlations $R(n)$.
Filled circles, empty circles, filled triangles and empty triangles for
one, two, three, and four holes doped to an $N=14$ orbital chain. 
}
\label{fig:rnjt}
\end{figure}

The energies of insulating and metallic phase are close to each other 
for typical parameters. Therefore, even a moderate JT energy 
$E_{\rm JT}=0.25t$ is sufficient to stabilize the insulating phase in a 
broad regime of doping, with occupied $|x\rangle$ orbitals fragmenting 
the chain into smaller units. When a single hole is doped, the JT field 
has qualitatively a similar effect as a larger value of $J$ --- a hole 
is then well localized within a three-site cluster, and the hole-orbital 
correlation $P(n)$ exhibits more pronounced alternation, indicating 
robust orbital order [Fig. \ref{fig:rnjt}(a)], as in Fig. 
\ref{fig:rn}(a). The orbital intersite correlations $R(n)$ are perfect 
already beyond the second nearest neighbor of the hole 
[Fig. \ref{fig:rnjt}(b)]. 

When more holes are added, the orbital order gradually softens, but a 
clear tendency towards orbital alternation is observed even for high 
doping with four holes [Fig. \ref{fig:rnjt}(b)]. Note that a second hole 
may be added at a fourth or sixth site away from the first one. This, 
together with weak hole hopping within the three-site clusters, weakens 
both hole-orbital correlations $P(n)$ and the orbital order $R(n)$ at 
$n=4$ and (less) at $n=6$. For the present case of $N=14$ sites these 
latter correlations are reduced more for $n\ge 4$ than for $n=2$. With a 
third hole added, we found separated holes in three-site clusters, with 
each hole occupying predominantly the central site of its cluster in 
order to minimize the JT energy. A typical inter-hole distance is then 
four lattice constants, and for this reason $P(4)$ becomes negative, 
and $R(4)$ is reduced stronger than $R(n)$ for any other $n>0$. The 
fourth hole is added at one of $|x\rangle$ orbitals within a longer
$|z\rangle-|x\rangle-|z\rangle-|x\rangle-|z\rangle$ unit, and makes then 
the first bigger cluster, with two holes and three mobile $|z\rangle$ 
electrons. Particularly in this regime of parameters and up to this 
doping regime, the distribution of holes is reminiscent of doping 
CuO$_3$ chains in YBa$_2$Cu$_3$O$_{6+x}$, where holes are doped first in 
separated CuO$_2$ units, where they are trapped, generating fragmented 
units of CuO$_3$ chains and causing jumps and plateaus for the hole 
counts in the CuO$_2$ planes as a function of doping.\cite{Zaa88}
 
\begin{figure}
\includegraphics[width=7.7cm]{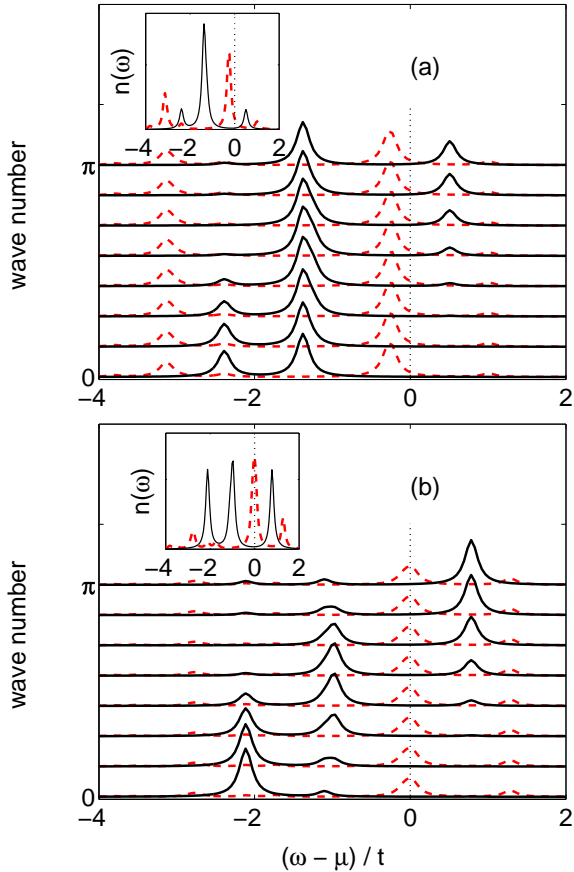}
\caption{(color online)
Spectral functions $A(k,\omega)$ for an insulating $N=14$ orbital chain 
for $J/t=0.125$ and $E_{\rm JT}=0.25t$, obtained for increasing doping 
by:
(a) one hole, and 
(b) three holes.  
Solid and dashed lines for $|z\rangle$ and $|x\rangle$ excitations. 
Insets and peak broadening as in Fig. \ref{fig:n=1}.
}
\label{fig:a125jt}
\end{figure}

The spectral functions $A(k,\omega)$, presented for two representative 
doping levels with one and three holes in Fig. \ref{fig:a125jt}, show 
that they are remarkably similar in the entire regime of doping $x<0.3$. 
Excitations in $|x\rangle$ orbitals lead to maxima below and above the 
Fermi energy, with the spectral weight moving gradually to higher 
energies under increasing doping. For hole excitations in $|z\rangle$ 
orbitals one finds three nondispersive features: the central peak for 
doping within the orbital ordered regions, and two side peaks at
energies corresponding to the excitations of orbital polarons. The 
spectral weight of the latter features increases on the cost of the 
central peak when doping increases, and the spectral weight distribution 
depends on momentum $k$. At doping increasing up to five holes (not 
shown) the spectra are still fairly localized, and the $|x\rangle$ 
spectral weight is further reduced for $\omega<\mu$. 

\begin{figure}
\includegraphics[width=7.7cm]{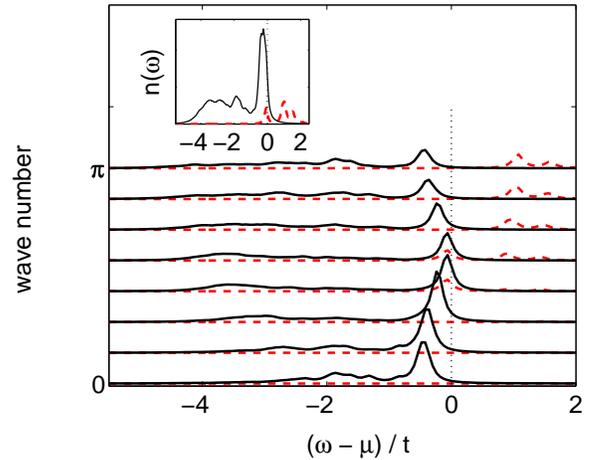}
\caption{(color online)
Spectral functions $A(k,\omega)$ for an $N=14$ spin $t$-$J$ chain 
(\ref{spintJ}) at doping by one hole, as obtained for $J/t=0.125$ with 
a staggered field $h_s=2J$. Solid and dashed lines indicate hole and 
electron excitations. Inset shows the density of states $n(\omega)$.
}
\label{fig:atj}
\end{figure}

A remarkable change of the spectral functions $A(k,\omega)$ has been 
found for the spin $t$-$J$ model in presence of a finite staggered field 
$h_s=2J$. The field generates a string potential, and simulates thus a 
2D model.\cite{Wro96} However, in this quantum model, in contrast to the 
orbital model shown in Fig. \ref{fig:a125jt}(a), a hole is not confined
within a polaron but can move coherently when it couples to quantum spin 
fluctuations, which repair the defects in the spin background generated
by a moving hole. As a result, a quasiparticle peak emerges close to the 
Fermi energy (Fig. \ref{fig:atj}), indicating a propagation of a hole 
dressed by spin excitations, with a dispersion $\sim 4J$ [note the units 
in Eq. (\ref{spintJ})]. The remaining spectral weight is distributed 
over incoherent features at lower energy while the low energy 
quasiparticle band visible in Fig.~\ref{fig:a1251h} has vanished as well 
as the incoherence of the band for small $k$. 

Moreover, the quasiparticle band has lost a simple tight-bindinglike 
dispersion found before at $h_s=0$, and instead of crossing the Fermi 
energy, it has periodicity $\pi$ because of the doubled unit cell and 
folds back toward lower energies after reaching the Fermi energy at 
$k=\pi/2$. The peaks are sharpest near this point, confirming the 
quasiparticle character of this excitation, but have lower spectral 
weight at $k=\pi$ than at $k=0$. The quasipartiple band close to the 
Fermi energy leads to a quite distinct peak in the density of states 
(see inset of Fig. \ref{fig:atj}). It is separated by a pseudogap from
the electronic excitations at $\omega>\mu$, and also by another 
pseudogap at $\sim 4J$ below the Fermi energy from the incoherent part 
of the spectrum at lower energies. Thus, the present case is radically 
different from the case of $h_s=0$ [see Fig. \ref{fig:n=1}(b)], and 
resembles the spectral functions for a 2D model.\cite{Mar91,Ram93} 

\section{Orbital polarons at finite temperature}
\label{sec:temp}

\subsection{Effective orbital $t$-$J$ model}
\label{sec:efftJ}

At increasing temperature the FM spin order assumed in Sec. 
\ref{sec:otj} is gradually destroyed and AF spin configurations on the 
bonds occur with finite probability, modifying the form of both $t$ and 
$J$ terms in Eq. (\ref{HtJ}). Even when the ground state at $T=0$ is FM, 
superexchange interactions which originate from $t_{2g}$ electron 
excitations play an important role and contribute at finite temperature. 
These interactions are frequently treated as an effective AF 
superexchange between core $S=3/2$ spins, but {\it de facto} they depend 
on the total number of $d$ electrons of two interacting Mn ions.
\cite{Ole02} We have verified, however, that the $t_{2g}$ superexchange 
terms derived for these different configurations are of the same order 
of magnitude, so it is acceptable to consider their effect as equivalent 
to a Heisenberg interaction with an average exchange constant $J'>0$. 
Therefore, we include in the present 1D model the spin interaction,
\begin{equation}
\label{HJ'}
H_{J'}=J'\sum_{i}\big({\vec S}_i\cdot{\vec S}_{i+1}-S^2\big). 
\end{equation}

The superexchange due to $e_{g}$ electron excitations contains spin 
scalar products multiplied by orbital interactions on the bonds, and 
the full many-body problem would require treating the coupled spin and 
orbital dynamics. Here we decouple spins and orbitals in the 
mean-field approximation, and study the orbital correlations and their 
consequences for the magnetic order by replacing the scalar products of 
spin operators on each bond by their average values,\cite{Dag02,Dag98} 
\begin{equation}
\label{uij}
\langle {\vec S}_i\cdot{\vec S}_{i+1}\rangle=S^2\big(2u_{i,i+1}^2-1\big), 
\end{equation}
where $u_{i,i+1}=\cos(\theta_{i,i+1}/2)e^{i\chi_{i,i+1}}$, and 
directions of two classical spins at sites $i$ and $i+1$ differ by angle 
$\theta_{i,i+1}$. The complex phase $\chi_{i,i+1}$ does not have any 
effect in the present 1D model.\cite{Kol02}

We investigate the {\it effective orbital $t$-$J$ model},
\begin{equation}
\label{HS}
{\cal H(S)}=H_t+H_J+H_{J'}+H_{\rm JT}, 
\end{equation}
where the AF interactions of core spins (\ref{HJ'}) and the JT field
acting on orbital variables (\ref{HJT}) are playing the role of external 
fields. The hopping term $H_t$ describes the dynamics of $e_g$ electrons 
which are locally coupled to $t_{2g}$ core spins by large Hund's 
exchange $J_H$ element, and have their spins parallel to them in the 
ground state. In agreement with double exchange mechanism,
\cite{Zen51,Bri99} the hopping along each bond $\langle i,i+1\rangle$ is 
then modified by the spin order to 
\begin{equation}
\label{ttilde}
{\tilde t}_{i,i+1}=tu_{i,i+1}, 
\end{equation}
and vanishes when the 
spins are antiparallel. This approximation allowed to establish the 
existence of magnetic polarons in the 1D Kondo model.\cite{Kol03}

The superexchange due to $e_g$ electrons $\propto J$ generates orbital 
interactions in $H_J$ which follow from virtual charge excitations, 
either for Mn$^{3+}$-Mn$^{3+}$, or for Mn$^{3+}$-Mn$^{4+}$ pairs. 
Consider first the spin-orbital model of Ref. \onlinecite{Fei99} for an 
undoped LaMnO$_3$. So far, we included the orbital superexchange [Eq.
(\ref{HJ}) for a situation when high-spin $e_g$ excitations dominate, 
but low-spin excitations would also contribute at finite temperature. 
Taking realistic parameters for the Coulomb interaction $U$ and Hund's 
exchange $J_H$, the low-spin $^4\!A_1$, $^4E$, and $^4\!A_2$ states have 
the energies: $7.3$, $7.8$, and $9.6$ eV.\cite{Fei99} We keep the 
relative importance of various excitations and write the terms resulting 
from various charge excitations to the superexchange, after averaging
them over the spin configuration, as follows:
\begin{eqnarray}
\label{app33}
\frac{1}{10}\;\frac{t^2}{\varepsilon(^6\!A_1)}\;
\langle {\vec S}_i\cdot{\vec S}_{i+1}\!+\!6\rangle 
\!&=&\!\frac{J}{5}(4u_{i,i+1}^2+1),           \nonumber \\
\frac{1}{16}\left[\frac{t^2}{\varepsilon(^4\!A_1)}\!+\!
       \frac{3}{5}\frac{t^2}{\varepsilon(^4E)}\right]
\langle {\vec S}_i\cdot{\vec S}_{i+1}\!-\!4\rangle \! &\simeq &\!
\frac{2J}{5}(u_{i,i+1}^2-1),               \nonumber \\  
\frac{1}{16}\left[\frac{t^2}{\varepsilon(^4E)}\!+\!
                  \frac{t^2}{\varepsilon(^4\!A_2)}\right]
\langle {\vec S}_i\cdot{\vec S}_{i+1}\!-\!4\rangle \!&\simeq &\!
\frac{9J}{20}(u_{i,i+1}^2-1),             \nonumber \\    
\end{eqnarray}

In a doped system one finds only AF superexchange terms for 
Mn$^{3+}$-Mn$^{4+}$ bonds [while FM ones are explicitly included as 
double exchange in the orbital model (\ref{HS})], which contribute only 
if an $e_g$ electron at site Mn$^{3+}$ occupies a directional orbital 
along the considered bond direction, in our case a $3z^2-r^2$ orbital,
and can hop to its Mn$^{4+}$ neighbor. Taking the excitation energy to 
the low-spin configuration $\varepsilon(^3E)\simeq 3$ eV,\cite{Ole02} 
one finds 
\begin{equation}
\label{app34}
\frac{1}{8}\,\frac{t^2}{\varepsilon(^3E)}\;
\langle {\vec S}_i\cdot{\vec S}_{i+1}-3\rangle\simeq J(u_{i,i+1}^2-1).
\end{equation}

Using the coefficients given in Eqs. (\ref{app33}) and (\ref{app34}),
one arrives at the effective orbital $t$-$J$ model (\ref{HS}), with
\begin{eqnarray}
\label{tT}
H_t&=&-\sum_{i}{\tilde t}_{i,i+1}
    \big({\tilde c}_{iz}^{\dagger}{\tilde c}_{i+1,z}^{}
        +{\tilde c}_{i+1,z}^{\dagger}{\tilde c}_{iz}^{}\big),        \\
\label{JT}
H_J&=& \frac{1}{5}J\sum_{i}\big(2u_{i,i+1}^2+3\big)
\Big( 2T_{i}^zT_{i+1}^z
-\textstyle{\frac{1}{2}{\tilde n}_i{\tilde n}_{i+1}} \Big) \nonumber \\
&-&\frac{9}{10}J\sum_{i}
\big(1-u_{i,i+1}^2\big){\tilde n}_{iz}{\tilde n}_{i+1,z}             \\
&-&J\sum_{i}\big(1-u_{i,i+1}^2\big)
\big[{\tilde n}_{iz}(1\!-\!{\tilde n}_{i+1})
+(1\!-\!{\tilde n}_i){\tilde n}_{i+1,z}\big].              \nonumber 
\end{eqnarray}
The hopping ($H_t$) and the interaction ($H_J$) terms depend on the 
actual spin configuration which fixes the bond variables $\{u_{i,i+1}\}$ 
[see Eq. (\ref{uij})]. In the FM state at $T=0$ all $u_{i,i+1}=1$, the
AF terms vanish, and one recovers the form of $H_t$ and $H_J$ used in 
Sec. \ref{sec:otj}. 

It is a crucial feature of the effective orbital model given by Eq.
(\ref{HS}) that spin interactions are influenced by orbital correlations 
along the chain, and the latter can support either FM (for 
alternating $x/z$ orbitals) or AF (for polarized $z$ orbitals) spin 
order. At $T=0$ one can find on optimal state by minimizing the total 
(internal) energy of the system, $E=\langle{\cal H(S)}\rangle$, over the 
spin and orbital configurations. In fact, at $T=0$ two solutions are 
possible, depending on the parameters in Eq. (\ref{HS}). Let us consider 
first a purely electronic model with $E_{JT}=0$ at half filling (we will 
show below that the situation is similar at $E_{JT}>0$, but the region 
of stability of the FM states is extended). If $J'=0$, FM order is 
stable and coexists with alternating orbital order. This situation was 
discussed in detail in Sec. \ref{sec:num}. However, a relatively small 
value of $J'=0.0125t$ is sufficient to compensate the energy difference 
between this state and an AF state with occupied $|z\rangle$ orbitals,
and (at $T=0$) one finds the latter state for $J'>0.0125t$. It is 
interesting to investigate a competition between these magnetic states 
at finite temperature when the chain is doped.

We investigated the spin and orbital correlations for the effective 
orbital $t$-$J$ model (\ref{HS}) at finite temperatures by employing a 
combination of a Markov chain Monte Carlo (MCMC) algorithm for the 
core spins, with Lanczos diagonalization for the many-body problem posed 
by the orbital chain of $N=12$ sites with periodic boundary conditions, 
at each given distribution of classical variables $\{u_{i,i+1}\}$ for 
$i=1,\cdots,12$, which stands for a particular spin configuration. 
The partition function to be then evaluated is 
\begin{equation}
  {\cal Z} = \int {\cal D}[{\cal S}]\; 
  \textrm{Tr}_c\, \textrm{e}^{-\beta {\cal H(S)}}\;,
\end{equation}
where $\beta=1/k_BT$ (we adopt the units with $k_B=1$), and
$\int {\cal D}[{\cal S}]$ denotes the integral over the $N$-dimensional 
space of all core spin configurations for the chain of length $N$. The 
core spins ${\cal S}\equiv\{{\vec S}_i\}$ determine the site-dependent 
hopping parameters $\{u_{i,i+1}\}$ and thus the fermionic Hamiltonian 
${\cal H(S)}$ is fixed. 

The trace over the fermionic degrees of freedom, 
$\textrm{Tr}_c\,\textrm{e}^{-\beta {\cal H(S)}}=:w({\cal S})$, gives 
the statistical weight for ${\cal S}$ and is sampled by the MCMC. For 
independent electrons it can easily be evaluated;\cite{Dag98,Fur98} for 
interacting electrons one has to use a Lanczos algorithm,\cite{Kol03} 
but as $w({\cal S})$ is strictly positive, one still has no sign 
problem. Since the position of the $|x\rangle$ electrons is conserved, 
the Hamiltonian (\ref{HS}) has a block-diagonal structure, with each 
block corresponding to one fixed distribution of $|x\rangle$ electrons 
along the chain. There are, of course, many such blocks, but their 
subsequent diagonalization is still much faster than the diagonalization 
of the complete matrix. For the extreme case of the completely filled 
chain, each block has dimension one, meaning that the Hamiltonian is 
already diagonal and no matrix-vector multiplication has to be 
performed. Furthermore, degenerate eigenvalues can be resolved, 
if they are in different blocks.

For the MCMC updates, $w({\cal S})$ was calculated from the lowest few 
eigenenergies of \emph{each} block for an $N=12$ chain, until a new 
Lanczos step did no longer modify the contribution from this block. 
Observables were only calculated from the lowest 14 eigenstates of the 
\emph{whole} space. In order to monitor this approximation, the 
Boltzmann factor of these states was measured. The weight of the highest 
included state was approximately $1.5$ percent for the worst case 
(filled chain, $\beta t=20$, $J'=0.02$, $E_{JT}=0$), $0.3-0.5$ percent 
for $\beta t=50$ and the filled chain and negligible for $\beta=100$ or 
finite doping. This means that for those observables, which are 
calculated from the eigenstates of the Hamiltonian, at most a few 
percent of the total weight were missed. For the MCMC updates and for
observables which do not need the eigenvectors (core spin correlation 
and total number of $|x\rangle/|z\rangle$ electrons), the error was 
even smaller.

The core spins were rotated for whole sections of the chain at once. 
Because acceptance for core spin updates was high, we performed two or 
three such rotations before testing acceptance. Every spin was therefore 
rotated several times per sweep. The numbers $N_x$ and $N_z$ of 
$|x\rangle$ and $|z\rangle$ electrons remained fixed for the evaluation 
of $w({\cal S})$, and every five updates we proposed to increase 
(decrease) $N_x$ and decrease (increase) $N_z$, respectively, thereby 
sampling $N_x$ and $N_z$. The total number of electrons $N_e$ was kept 
fixed. Between measurements, $40$ to $100$ lattice sweeps (depending on
temperature) were done in order to decorrelate the samples. 
We then employed autocorrelation analysis and found the samples to be 
uncorrelated. To reduce the statistical errors, 200 uncorrelated samples 
were obtained for each set of parameter values.

\subsection{Orbital order versus spin order}
\label{sec:ct}

\begin{figure}
\includegraphics[width=7.7cm]{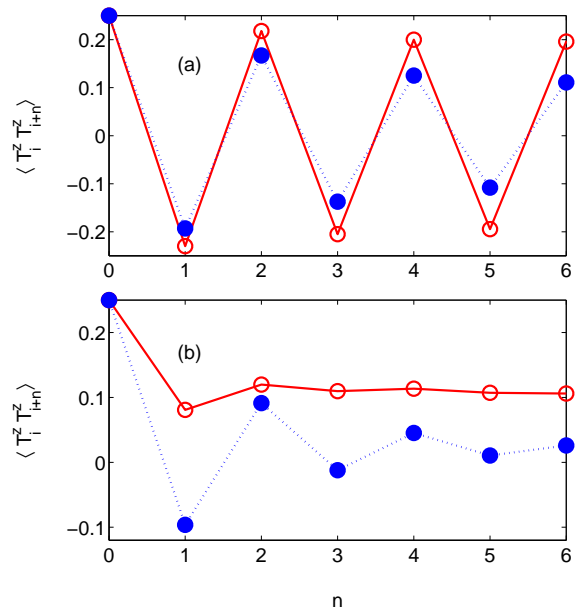}
\caption{
Orbital correlations $\langle T_i^zT_{i+1}^z\rangle$ ($\ref{oo}$) 
at half filling, as obtained at low ($\beta t=100$, open circles), 
and at intermediate ($\beta t=50$, filled circles) temperature 
for an $N=12$ site orbital chain (\ref{HS}), with $J=0.125t$, and:
(a) $J'=E_{\rm JT}=0$,
(b) $J'=0.02t$, $E_{\rm JT}=0$, and 
(c) $J'=0.02t$, $E_{\rm JT}=0.25t$. 
Statistical errors are smaller than the symbol sizes.
}
\label{fig:tt}
\end{figure}

In addition to the correlation functions studied in Sec. \ref{sec:num}, 
spin correlations $\langle S_i^zS_j^z\rangle$ are now investigated by 
evaluating the spin structure factor, 
\begin{equation}
\label{s(k)}
S(k)=\frac{1}{N^2}\sum_{ij}e^{ik(R_i-R_j)}\langle S_i^zS_j^z\rangle,          
\end{equation}
which depends on the 1D momentum $k$, and follows from Monte-Carlo 
simulations of $N=12$ chains. The averages 
$\{\langle S_i^zS_{i\pm 1}^z\rangle\}$ are calculated from the spin 
configurations ${\cal S}\equiv\{{\vec S}_i\}$  determined by the MCMC 
updates for $\{u_{i,i\pm 1}\}$.

\begin{figure}
\includegraphics[width=7.7cm]{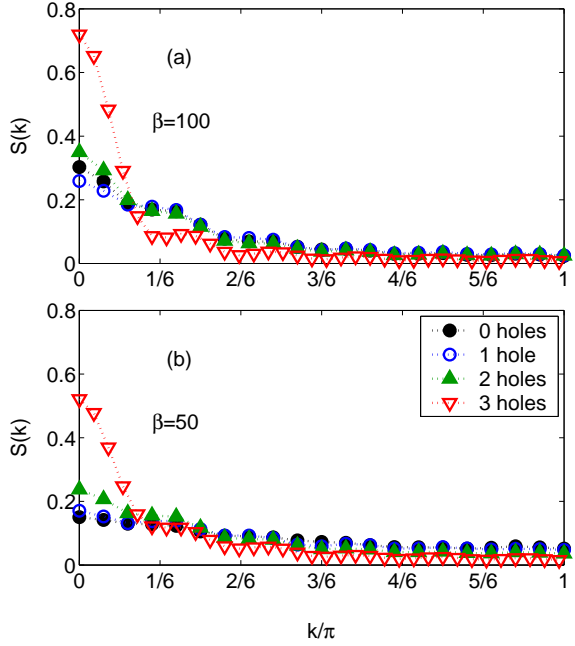}
\caption{(color online)
Spin structure factor $S(k)$ as obtained for an undoped $N=12$ orbital 
chain (\ref{HS}), and for increasing hole doping up to five holes: 
(a) at low temperature $\beta=100$, and 
(b) at intermediate temperature $\beta=50$ (in units of $t^{-1})$.
Statistical errors are smaller than the symbol sizes.
Parameters: $J=0.125t$, $J'=0$, $E_{\rm JT}=0$. 
}
\label{fig:sk0}
\end{figure}

The interplay between spin and orbital correlations becomes transparent
by varying the core spin AF superexchange $J'$ (\ref{HJ'}) due to 
$t_{2g}$ electrons. We investigated spin and orbital order for two 
characteristic values of temperature: $\beta t=100$ and $\beta t=50$,
corresponding to $T\sim 60$ and $\sim 120$ K for $t\sim 0.5$ eV, i.e.,
well below the magnetic transition. First we consider the case of $J'=0$
which reproduces the ground state analyzed in Secs. \ref{sec:otj} and 
\ref{sec:num}. While the orbital alternation, measured by $T(n)$ 
correlation function (\ref{oo}), is perfect at $T=0$, it softens 
somewhat when temperature increases, but is still robust at temperature 
$\beta t=50$, as shown in Fig. \ref{fig:tt}(a). Orbital alternation 
supports the FM spin order at half filling, which gives a distinct 
maximum at $k=0$ of the spin structure factor $S(k)$ (see Fig. 
\ref{fig:sk0}). In the weak doping regime with one or two holes added, 
the spin correlations, $\langle n_iS_{i+n}^zS_{i+n+1}^z\rangle$, 
are driven by superexchange and are FM at any distance $n$ from the 
hole, and only weakly depend on $n$. These correlations increase by a 
factor close to two when the doping changes from two to three holes. 
This explains why the maximum of $S(k)$ at $k=0$ remains almost 
unchanged in the low doping regime by one or two holes, but is next 
strongly enhanced when doping increases to three holes. Precisely at 
this concentration electrons redistribute within the chain and occupy 
practically only $|z\rangle$ orbitals, giving a metallic state. This 
demonstrates that double exchange plays a primary role in the observed 
insulator-metal transition and significantly enhances the stability of 
the FM order in the metallic phase. 

\begin{figure}
\includegraphics[width=7.7cm]{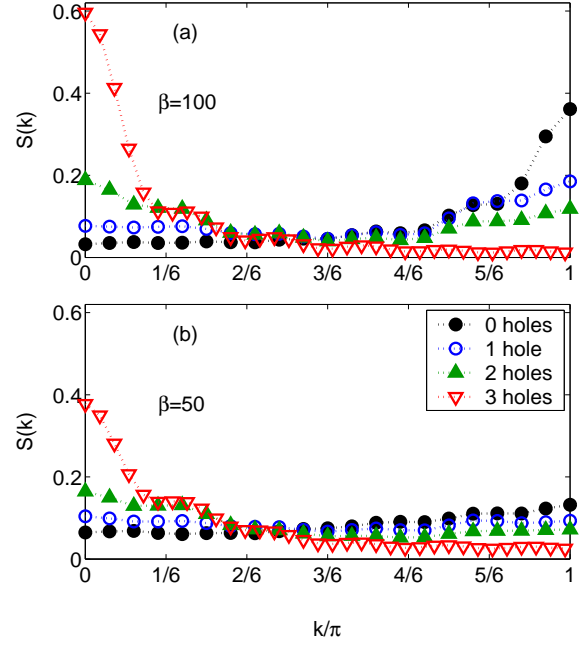}
\caption{(color online)
Spin structure factor $S(k)$ for an $N=12$ orbital chain (\ref{HS}) as 
in Fig. \ref{fig:sk0}, but for $J'=0.02t$. 
}
\label{fig:sk02}
\end{figure}

At $J'=0.02t$ the orbital correlations found in the half-filled 
chain are markedly different [Fig. \ref{fig:tt}(b)]. At low temperature 
($\beta t=100$) they indicate that primarily (but not only) $|z\rangle$ 
orbitals are occupied, while no $|x\rangle$ electrons were found in the 
ground state at $T=0$. This orbital state is induced by finite $J'$, 
and supports the AF spin interactions due to $e_g$ excitations which 
select then low-spin states in the $e_g$ excitations along 
$|z\rangle-|z\rangle$ bonds. When a single hole is doped, 
the spin correlations remain still AF at low temperature [Fig. 
\ref{fig:sk02}(a)], except for the spin-spin correlation between the 
site occupied by the hole and its nearest neighbor, giving rise to a 
small FM polaron. Therefore, the value of $S(0)$ increases already at 
this low doping, and the electron density in $|z\rangle$ orbitals is 
enhanced close to the hole itself. 

When temperature increases, it becomes clear that the orbital 
$|x\rangle/|z\rangle$ alternation, supporting the FM spin interactions,
competes at half filling with the above uniformly polarized chain with 
occupied $|z\rangle$ orbitals, supporting the AF spin order. A clear 
tendency towards orbital alternation is detected by a negative 
nearest-neighbor orbital correlation $T(1)\simeq -0.10$ [Fig. 
\ref{fig:tt}(b)] at higher temperature ($\beta t=50$). Although the 
value of $S(\pi)$ is still larger than $S(0)$ at half filling, both 
(weak) maxima, corresponding to FM and AF order, become almost equal 
already for doping with one hole [Fig. \ref{fig:sk02}(b)]. This case is 
qualitatively similar to a chain doped by two holes at $\beta t=100$ 
[Fig. \ref{fig:sk02}(a)], where also two (stronger) maxima of $S(k)$ 
indicate coexisting islands of FM and AF spin correlations along the 
chain. Thus, we found that the AF correlations are gradually changed 
into FM ones with increasing temperature. This trend follows from the 
difference in the energy scales --- when the thermal magnetic 
excitations destroy the energy gains due to $J$ and $J'$, the kinetic 
energy $\propto t$ is a dominating energy which can be optimized by 
selecting $|z\rangle$ occupied orbitals and FM spin correlations. 
 
At doping by three holes the spin correlations are FM, both at low 
($\beta t=100$) and at higher temperature ($\beta t=50$), showing that 
the double exchange enhances the effective FM interactions and changes 
the characteristic temperature at which the spin correlations weaken 
--- thus the Curie temperature $T_C$ would increase in a 3D case. This 
agrees with the experimental observations --- indeed, the Curie 
temperature increases with hole doping $x$ in the metallic regime.
\cite{End97,Bio01} 
 
\begin{table}
\caption{
Total number of electrons in itinerant orbitals $N_z$ (top part) and in
localized orbitals $N_x$ (bottom part) for different doping level $x$ 
and for increasing temperature $\beta t$, as obtained in the Monte-Carlo 
simulations for an $N=12$ orbital chain [Eq. (\ref{HS})] (statistical 
errors are also given). 
Parameters: $J=0.125t$, $J'=0.02t$, $E_{\rm JT}=0$.
}
\vskip .3cm
\begin{ruledtabular}
\begin{tabular}{ccccc}
 & \multicolumn{4}{c}{$N_z$} \cr
 $\beta t$   &  $x=0$ &  $x=1/12$ &  $x=1/6$ &  $x=1/4$  \cr
\colrule
 100  & $9.97\pm 0.09$ & $9.80\pm 0.06$ & $9.64\pm 0.04$ & 
                                          $9.0 \pm 0.0 $ \cr
  50  & $7.85\pm 0.07$ & $8.51\pm 0.06$ & $8.81\pm 0.04$ & 
                                          $8.98\pm 0.01$ \cr
  30  & $7.47\pm 0.07$ & $8.05\pm 0.06$ & $8.53\pm 0.05$ & 
                                          $8.88\pm 0.03$ \cr
  20  & $7.34\pm 0.07$ & $7.85\pm 0.06$ & $8.18\pm 0.06$ & 
                                          $8.55\pm 0.04$ \cr
\colrule
 & \multicolumn{4}{c}{$N_x$} \cr
 $\beta t$   &  $x=0$ &  $x=1/12$ &  $x=1/6$ &  $x=1/4$  \cr
\colrule
 100  & $2.03\pm 0.09$ & $1.20\pm 0.06$ & $0.36\pm 0.04$ & 
                                          $0.0 \pm 0.0 $ \cr
  50  & $4.15\pm 0.07$ & $2.49\pm 0.06$ & $1.19\pm 0.04$ & 
                                          $0.02\pm 0.01$ \cr
  30  & $4.53\pm 0.07$ & $2.95\pm 0.06$ & $1.47\pm 0.05$ & 
                                          $0.12\pm 0.03$ \cr
  20  & $4.66\pm 0.07$ & $3.15\pm 0.06$ & $1.82\pm 0.06$ & 
                                          $0.45\pm 0.04$ \cr
\end{tabular}
\end{ruledtabular}
\label{tab:nznx}
\end{table}

Although at $T=0$ one finds indeed the AF ground state with $N_z=12$ for 
a half-filled $N=12$ chain, many excited states with a few $|x\rangle$ 
orbitals occupied are found at low energy and contribute already at 
$\beta t=100$. As a result, electrons are redistributed over $e_g$ 
orbitals by thermal excitations, and the orbital polarization at 
$\beta t=100$ is far from complete, with $N_z\simeq 10$ and 
$N_x\simeq 2$ (Table \ref{tab:nznx}). When temperature increases further
to $\beta t=50$ ($\beta t=20$), orbital disorder increases and one finds 
$N_z\simeq 7.8$ and $N_x\simeq 4.2$ ($N_z\simeq 7.3$ and 
$N_x\simeq 4.7$). This demonstrates that the balance between FM and AF 
terms realized at half filling is rather subtle --- increasing 
temperature favors more disorder in the chain which destroys a uniform 
AF phase, supported by charge excitations along $|z\rangle-|z\rangle$ 
bonds.

The population of localized $|x\rangle$ states in the chain decreases 
quite fast with doping, particularly in the regime of $\beta t\leq 100$. 
At $x=1/4$ one finds almost no $|x\rangle$ defects in the metallic 
chains, except at rather high temperature $\beta t=20$. Increasing 
electron density in $|x\rangle$ orbitals with increasing temperature 
may be seen as a precursor effect for the metal-insulator transition at 
$T_C$ in the intermediate doping regime.   

\subsection{Polaronic features at finite doping}
\label{sec:spectrat}

The spectral functions $A(k,\omega)$ obtained for the FM phase at half 
filling for $J'=0$ (not shown) are similar to those discussed in Sec. 
\ref{sec:spectra}. Apart from some broadening due to finite $T$, the 
one-hole excitations are again fairly localized, similar to those shown 
in Fig. \ref{fig:n=1}. In the doped regime one finds first the localized
spectra at low doping, similar to those of Fig. \ref{fig:a1251h}(a), 
with a distint pseudogap at the Fermi energy $\mu$. As found before at 
$T=0$, doping by three holes suffices for the crossover to the metallic 
phase, with somewhat broadened peaks in $A(k,\omega)$, following the 
one-particle dispersion due to the hopping within $|z\rangle$ orbitals, 
like in Fig. \ref{fig:a1253h}(a).

\begin{figure}
\includegraphics[width=7.7cm]{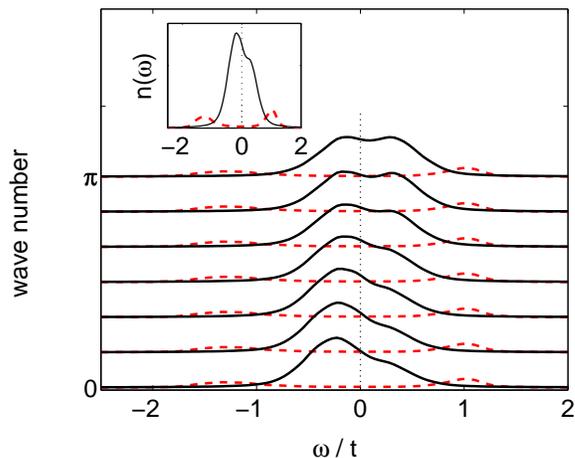}
\caption{(color online)
Spectral functions $A(k,\omega)$ for a half-filled $N=12$ orbital chain 
(\ref{HS}) at temperature $\beta t=100$.
Solid and dashed lines for $|z\rangle$ and $|x\rangle$ excitations. 
Inset shows the density of states $n(\omega)$.
Parameters: $J=0.125t$, $J'=0.02t$, $E_{\rm JT}=0$. 
}
\label{fig:sfh0}
\end{figure}

In contrast, the spectral properties obtained for the AF phase found at 
half filling with $J'=0.02t$ are quite different (Fig. \ref{fig:sfh0}). 
First of all, there are predominantly $|z\rangle$ electrons at low 
temperature $\beta t=50$, and still more so at $\beta t=100$, so the 
spectral weight of the localized $|x\rangle$ hole excitations, found at 
energies $\omega\sim -1.3t$ and $\sim t$ [Fig. \ref{fig:tt}(b)], appears 
to be very low. The $|z\rangle$ excitations have insulating character 
and are incoherent. They are given by a superposition of two features, 
namely a dispersionless peak at the hole binding energy $\omega\sim -2J$ 
and a weakly dispersive band with bandwidth $\sim 0.5t$. The first 
feature stems from the few $|z\rangle$ electrons which are surrounded by 
$|x\rangle$ electrons and is similar to that discussed in Sec. 
\ref{sec:undoped}. The second feature comes from electrons moving within 
the $|z\rangle$ polarized parts, with the strongly renormalized hopping 
$\propto{\tilde t}_{i,i+1}$ (\ref{ttilde}) for aligned $|z\rangle$ 
orbitals along an almost perfect AF bond. If the 
AF order were perfect, the bandwidth would vanish, because no hopping 
would then be possible. Both the dispersive and the dispersionless 
structures are broadened owing to thermal fluctuations of the core spins.
The superposition of these two bands yields a single peak with large
intensity  for $k=0$, because the bandwidth is approximately 
$\sim 0.5t=4J$ and the both bands therefore coincide at this point. For 
the same reason, the spectral weight at $k=\pi$ consists of two almost 
symmetric maxima. This interpretation is corroborated by the results for 
$\beta t=50$ (not shown), where (i) the dispersionless feature has a 
higher weight, because the population of $|x\rangle$ orbitals increases 
with increasing temperature (see Table I), and (ii) the structures in 
$A(k,\omega)$ are still more broadened because of the larger thermal 
fluctuations.

\begin{figure}
\includegraphics[width=7.7cm]{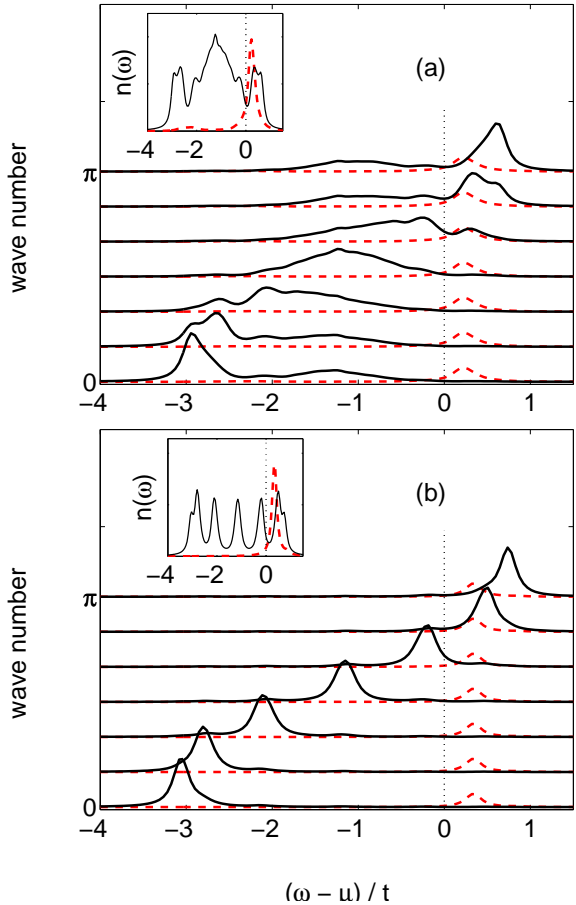}
\caption{(color online)
Spectral functions $A(k,\omega)$ for an $N=12$ orbital chain (\ref{HS})
at temperature $\beta t=100$ doped by: 
(a) two holes, and 
(b) three holes.
Solid and dashed lines, inset and parameters as in Fig. \ref{fig:sfh0}.
}
\label{fig:sfh23}
\end{figure}

The spectrum changes rapidly when doping increases. At doping of 
$x=1/12$ (one hole in an $N=12$ chain) a polaronic peak is found above 
$\mu$, while below $\mu$ the spectrum separates into a broad incoherent 
part at intermediate energies, and a low-energy peak with large 
intensity for low values of $k=0$ and $k=\pi/6$, being a symmetric image 
of the polaronic peak (not shown). As $N_x$ decreases (Table I), the 
$|x\rangle$ intensity below $\mu$ drops in doped chains. At doping 
$x=1/6$, $|x\rangle$ electrons were found only in some samples at low
temperature ($\beta t=100$). It seems that these fluctuations alone  
would not suffice to make the system insulating, but AF correlations are
here still strong (Fig. \ref{fig:sk02}) and generate a well-formed 
polaronic peak in the density of states for the electronic exciations
(with $\omega>\mu$), separated by a pseudogap from the incoherent 
spectrum due to $|z\rangle$ hole excitations [Fig. \ref{fig:sfh23}(a)]. 
In this case the (localized) $|x\rangle$ excitations are predominantly 
electronic, leading to a distinct nondispersive peak above $\mu$.  
Except for these latter excitations, the situation is here very similar 
to that found in the one-orbital model,\cite{Kol03} where FM polarons, 
generated dy doping, were embedded into an AF background.

\begin{figure}
\includegraphics[width=7.7cm]{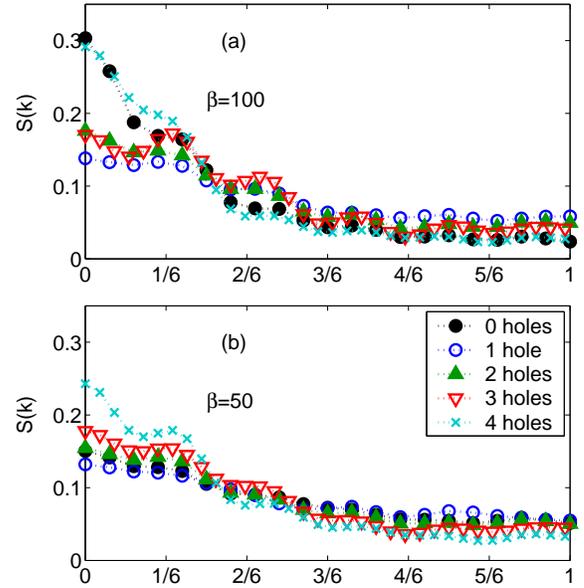}
\caption{(color online)
Spin structure factor $S(k)$ as in Fig. \ref{fig:sk0}, but for up to 
four holes. Parameters: $J=0.125t$, $J'=0.02t$, $E_{\rm JT}=0.25t$. 
}
\label{fig:sk0225}
\end{figure}

For the present parameters, doping $x=1/4$ (by three holes) gives an 
almost perfect metallic chain (occupied $|z\rangle$ orbitals) at low 
temperature $\beta t=100$, and the $|z\rangle$ excitations are then
coherent [Fig. \ref{fig:sfh23}(b)]. One finds seven excitations 
obtained for different momenta which sum up to five distinct structures 
in the density of states $n(\omega)$ --- the two side features consist 
of two peaks each, representing joined intensities for $k=0,\pi/6$ and 
for $k=5\pi/6,\pi$, respectively. An apparent pseudogap at $\omega=\mu$ 
is a finite size effect and would disappear at large system sizes.  

\subsection{Jahn-Teller effect at finite temperature}
\label{sec:JTt}

We have shown above that the orbital and magnetic order are interrelated
and influence each other. Therefore, as pointed out before,\cite{Fei99} 
not only the superexchange, but also purely orbital interactions which 
follow from the JT effect are of importance for the observed magnetic 
$A$-AF order in LaMnO$_3$. Now we will show that also in the present 1D 
model the JT effect may modify the magnetic order. This is particularly
transparent by considering a uniformly polarized insulating state with
occupied $|z\rangle$ orbitals, coexisting with an AF order, and 
stabilized by weak superexchange $J'=0.02t$ (Fig. \ref{fig:sk02}). If 
an alternating JT potential, given by Eq. (\ref{HJT}), increases, the 
AF order is easily destabilized --- already at $E_{\rm JT}=0.25t$ we 
found an almost perfect orbital staggering in the broad temperature 
regime at half filling [Fig. \ref{fig:tt}(c)], which induces instead the 
FM spin order, visible as a broad maximum in $S(k)$ centered at $k=0$ 
(Fig. \ref{fig:sk0225}). However, this maximum is less pronounced and 
other correlations $S(k)$ with $k>0$ are also present, unlike for 
$J'=E_{\rm JT}=0$, showing that the FM order induced by the JT potential
is definitely much weaker. 

\begin{figure}
\includegraphics[width=7.7cm]{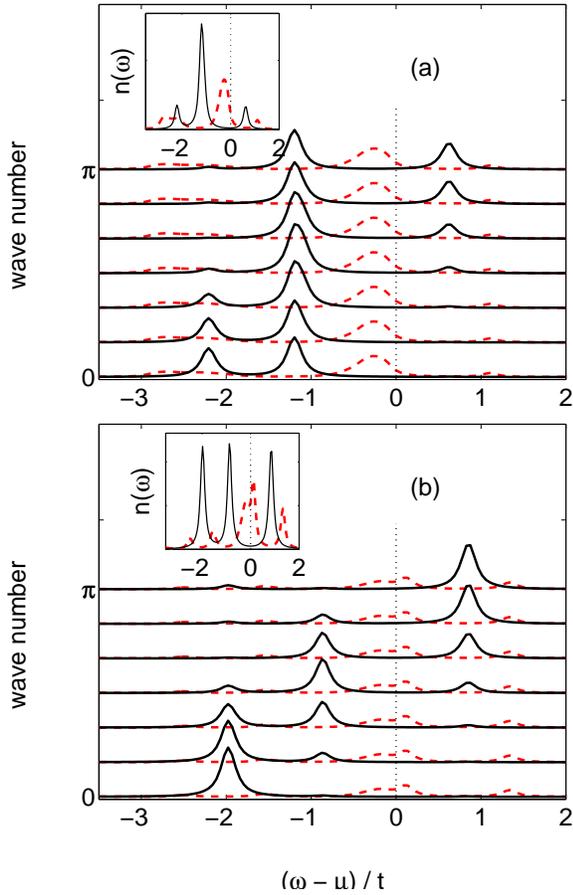}
\caption{(color online)
Spectral functions $A(k,\omega)$ for an $N=12$ orbital chain (\ref{HS}),
as obtained at temperature $\beta t=100$ for: 
(a) three holes, and 
(b) five holes.
Solid/dashed lines and inset as in Fig. \ref{fig:sfh0}.
Parameters: $J=0.125t$, $J'=0.02t$, $E_{\rm JT}=0.25t$.
}
\label{fig:sfh35}
\end{figure}

Even for moderate JT potential $E_{\rm JT}=0.25t$, increasing doping 
does not change the spectral properties qualitatively, and they remain 
dominated by localized excitations. For doping with one hole (not 
shown), two localized $|x\rangle$ excitations are accompanied by three 
localized $|z\rangle$ excitations: the central peak corresponding to a 
localized hole between two $|x\rangle$ electrons, and two satellite 
structures, a hole excitation at energy $\omega-\mu\sim -2t$ (with large 
intensity for low $k$ values), and an electron excitation at energy 
$\omega-\mu\sim 0.5t$, mainly contributing for large $k$ values. These 
structures are similar to those found before at $T=0$ [Fig. 
\ref{fig:a125jt}(a)], but they are now broadened by thermal spin 
fluctuations. When doping increases, the intensity of $|x\rangle$ hole 
excitations decreases and moves to the Fermi energy $\omega=\mu$, while 
a satellite peak grows for electron excitations ($\omega>\mu$), as shown 
for doping $x=1/4$ in Fig. \ref{fig:sfh35}(a). At this doping the FM 
correlations are enhanced by a factor $\sim 1.5$ with respect to the 
undoped case, but are still much weaker than those found before at the 
same doping for $E_{\rm JT}=0$ (Fig. \ref{fig:sk02}), where the FM phase 
was metallic. This demonstrates that the intersite magnetic correlations 
and the energy responsible for them are weaker in insulating FM 
manganites than in the metallic ones at the same doping. A similar trend 
was indeed observed for the values of the Curie temperature $T_C$,
\cite{Bac98} and one expects that also spin stiffness should be reduced 
by the JT coupling in CMR manganites.\cite{Mae03} 

The transition to metallic phase is damped by the JT distortions, and 
although the FM correlations increase significantly at doping $x=1/3$ 
(Fig. \ref{fig:sk0225}), the chain remains insulating. Only for as high 
doping as $x=5/12$, the $|x\rangle$ electrons are practically eliminated 
in the considered temperature range $\beta t\geq 50$, and a metallic 
behavior takes over. This metallic state gives almost free dispersion of 
$|z\rangle$ states in $A(k,\omega)$, while $|x\rangle$ (electron) 
excitations are again localized [Fig. \ref{fig:sfh35}(b)].

\section{Summary and conclusions}
\label{sec:sum}

The present study clarifies that orbital degrees of freedom are of 
crucial importance for the understanding of magnetic correlations in CMR 
manganites. First of all, for the realistic parameters of manganites 
the FM and AF state are nearly degenerate at half filling. In both cases 
the decisive term stabilizing the magnetic order originates from the 
$e_g$ superexchange. Although the ground state at $T=0$ would be FM 
in the absence of the AF superexchange between core spins, is is easy to 
flip the balance of magnetic (and orbital) interactions and stabilize 
instead the AF order in a purely electronic model. Here we adopted the 
AF interaction $J'=0.02t$ between core spins in order to stabilize the 
AF spin order in an undoped chain (the true value of $J'\simeq 0.004t$ 
in LaMnO$_3$, estimated from the value of N\'eel temperature of 
CaMnO$_3$, is smaller by a factor close to five\cite{Fei99}), and to 
demonstrate a gradual crossover from an AF insulator to a FM metal under 
increasing doping. In this regime of parameters the present 1D chain 
stands for the AF order along $c$ axis in the $A$-AF phase realized in 
LaMnO$_3$. 

We have shown that {\it even in the 1D model\/} the magnetic 
interactions are internally frustrated, with competing FM and AF terms 
in the superexchange. A delicate balance between these terms is easily
disturbed by the JT potential originating from the lattice. This allows 
one to investigate within the same framework both types of magnetic 
order which coexist in the $A$-AF phase of undoped manganites. On one 
hand, if the oxygen distortion have almost no influence, or if a uniform 
polarization in $|z\rangle$ orbitals would be induced by them, the AF 
spin order, found along $c$ axis in $A$-AF phase of LaMnO$_3$, would be
supported. On the other hand, if the oxygen distortions are 
alternating, they induce alternating orbital order, as it happens in 
$(a,b)$ planes of LaMnO$_3$, and the FM terms are selected from the 
$e_g$ superexchange. Note, however, that the FM correlations are weakly 
reduced from their values found in the absence of the JT interations 
($J'=0$ and $E_{\rm JT}=0$), demonstrating that the FM interactions do 
not depend {\it explicitly\/} on the JT terms, but are only induced by 
a given type of orbital order. In this way the JT coupling to the 
lattice helps to remove the frustration of magnetic interactions in
CMR manganites. 

The evolution of spin correlations for increasing doping showed that 
indeed two mechanism are responsible for ferromagnetism: when the
weakly doped 1D chain is insulating, the FM interactions are induced 
both by the {\it superexchange\/} terms following from the high-spin 
excited states, and by the local double exchange within polaronic 
states around single holes trapped in the insulating phase. The {\it 
double exchange\/} interactions is much stronger than the superexchange, 
and it fully takes over and operates in the metallic phase at higher 
doping. The difference between these two mechanisms is reflected by a 
fast increase of FM correlations at the insulator-metal transition which 
was investigated both within a purely electronic model, and including 
the JT potential induced by the lattice. 

It is quite remarkable that {\it orbital polarons\/} found in the 
present model with orbital degeneracy in the regime of AF spin 
correlations resemble FM polarons which occur in the Kondo model.
\cite{Kol03} This provides some support to a simplified picture of a 
nondegenerate conduction band which is able to capture the essential 
physics when the orbitals are polarized, and the orbital degrees of 
freedom are quenched and do not contribute in any significant way. A 
conservative point of view, based on double exchange mechanism, is that 
the FM polarons compete with the AF order and cause a transition to the 
metallic FM phase. Yet, this is not the only possibility --- we have 
shown that the FM phase at low doping could be insulating due to 
immobile orbital polarons, which allow to understand why this phase 
could be FM and insulating at the same time. Such polarons are expected 
to play an essential role in the insulator-metal transition within the 
FM phase in manganites.  
  
We believe that many qualitative features found in the present 1D study 
are generic for the interplay between orbital and magnetic order in CMR
manganites. Work is in progress on higher dimensional systems. Among 
others, an interesting question is to what extent the orbital order is 
modified when two $e_g$ orbitals start to fluctuate more strongly in 
either 2D or 3D systems, both due to quantum effects and due to 
increasing doping.

\begin{acknowledgments}
We thank E. Arrigoni and A. Pr\"ull for valuable discussions. 
This work has been supported by the Austrian Science Fund (FWF), 
Project No.~P15834-PHY.
A.~M.~Ole\'s would like to acknowledge the kind hospitality of Institute 
of Theoretical and Computational Physics, Graz University of Technology, 
and support by the Polish State Committee of Scientific Research (KBN) 
under Project No.~1 P03B 068 26.
\end{acknowledgments}



\begin{thebibliography}{00}

\bibitem{Dag02} E. Dagotto,
                   {\it Nanoscale Phase Separation in Manganites} 
                   (Springer-Verlag, Heidelberg, 2002);
                E. Dagotto, T.~Hotta, and A. Moreo,
                   Phys. Rep. {\bf 344}, 1 (2001).

\bibitem{Sch95} S. Jin, T. H. Tiefel, M. McCormack, R. A. Fastnacht,
                   R. Ramesh, and L. H. Chen,
                   Science {\bf 264}, 413 (1994);
                P. Schiffer, A. P. Ramirez, W. Bao, and S.-W. Cheong,
                   \prl {\bf 75}, 3336 (1995);
                A. Urushibara, Y.~Moritomo, T. Arima, A. Asamitsu, 
                   G. Kido, and Y.~Tokura,
                   \prb {\bf 51}, 14103 (1995).
                   
\bibitem{Zen51} C. Zener,
                   Phys. Rev. {\bf 82}, 403 (1951).

\bibitem{Ram04} T. V. Ramakrishnan, H. R. Krishnamurthy, S. R. Hassan, 
                   and G. Venketeswara Pai,
                   \prl {\bf 92}, 157203 (2004).

\bibitem{Dag98} S. Yunoki, J. Hu, A. L Malvezzi, A. Moreo, 
                   N. Furukawa, and E. Dagotto,
                   \prl {\bf 80}, 845 (1998);
                E.~Dagotto, S. Yunoki, A. L Malvezzi, A. Moreo, J. Hu,
                   S. Capponi, D. Poilblanc, and  N. Furukawa,
                   \prb {\bf 58}, 6414 (1998).

\bibitem{Fur98} N. Furukawa,
                   in: {\it Physics of Manganites} 
                   (Kluwer Academic Publisher, New York, 1998).

\bibitem{Edw99} D. M. Edwards, A. C. Green, and K. Kubo, 
                   J. Phys.: Condens. Matter {\bf 11}, 2791 (1999);
                A. C. Green and D.~M.~Edwards,
                   {\it ibid.\/} {\bf 11}, 10511 (1999); 
                M. Hohenadler and D. M. Edwards,     
                   {\it ibid.\/} {\bf 14}, 2547 (2002). 

\bibitem{Ali01} H. Aliaga, B. Normand, K. Hallberg, M. Avignon, 
                   and B. Alascio,
                   \prb {\bf 64}, 024422 (2001).

\bibitem{San02} W. Nolting, G. G. Reddy, A. Ramakanth, and D. Meyer, 
                   \prb {\bf 64}, 155109 (2001);
                C. Santos and W. Nolting,
                   {\it ibid.\/} {\bf 66}, 019901 (2002).
                    
\bibitem{Kol02} W. Koller, A. Pr\"ull, H. G. Evertz, 
                   and W. von der Linden,
                   \prb {\bf 66}, 144425 (2002);
                   {\it ibid.\/} {\bf 67}, 104432 (2003).
                    
\bibitem{Kol03} W. Koller, A. Pr\"ull, H. G. Evertz, 
                   and W. von der Linden,
                   \prb {\bf 67}, 174418 (2003).

\bibitem{Hor99} P. Horsch, J. Jakli\v{c}, and F. Mack,
                   \prb {\bf 59}, 6217 (1999);
                J. Ba\l{}a, P. Horsch, and F. Mack,
                   {\it ibid.\/} {\bf 69}, 094415 (2004).

\bibitem{Bri99} J. van den Brink and D. I. Khomskii,
                   \prl {\bf 82}, 1016 (1999);
                K. Held and D. Vollhardt,   
                   {\it ibid.\/} {\bf 84}, 5168 (2000);
                M. S. Laad, L. Craco, and E. M\"uller-Hartmann,
                   \prb {\bf 63}, 214419 (2001).

\bibitem{Ole02} A. M. Ole\'s and L. F. Feiner,
                   \prb {\bf 65}, 052414 (2002).

\bibitem{Kug82} K. I. Kugel and D. I. Khomskii,
                   Usp. Fiz. Nauk {\bf 136}, 621 (1982)
                   [Sov. Phys. Usp. {\bf 25}, 231 (1982)].

\bibitem{Tok00} Y. Tokura and N. Nagaosa,
                   Science {\bf 288}, 462 (2000);
                A.~M.~Ole\'s,
                   Phys. Stat. Sol. (b) {\bf 236}, 281 (2003).  

\bibitem{Fei97} L. F. Feiner, A. M. Ole\'s, and J. Zaanen,
                   \prl {\bf 78}, 2799 (1997);
                A. M. Ole\'s, L. F. Feiner, and J. Zaanen,
                   \prb {\bf 61}, 6257 (2000).
                   
\bibitem{Shi97} R. Shiina, T. Nishitani, and H. Shiba,
                   J. Phys. Soc. Jpn. {\bf 66}, 3159 (1997).
                           
\bibitem{Mae98} S. Maezono, S. Ishihara, and N. Nagaosa,            
                   \prb {\bf 57}, R13993 (1998);
                   {\it ibid.\/} {\bf 58}, 11583 (1998).

\bibitem{Fei99} L. F. Feiner and A. M. Ole\'s,
                   \prb {\bf 59}, 3295 (1999).

\bibitem{Oka02} S. Okamoto, S. Ishihara, and S. Maekawa, 
                   \prb {\bf 65}, 144403 (2002).
                    
\bibitem{Rod98} Y. Murakami, J. P. Hill, D. Gibbs, M. Blume, I. Koyama, 
                   M. Tanaka, H. Kawata, T. Arima, Y. Tokura, K. Hirota, 
                   and Y. Endoh,
                   \prl {\bf 81}, 582 (1998);
                J. Rodr\'{i}guez-Carvajal, M. Hennion, F. Moussa,
                   A.~H.~Moudden, L. Pinsard, and A. Revcolevschi,
                   \prb {\bf 57}, R3189 (1998).

\bibitem{vdB99} J. van den Brink, P. Horsch, F. Mack, and A. M. Ole\'s,
                   \prb {\bf 59}, 6795 (1999).

\bibitem{Bio01} G. Biotteau, M. Hennion, F. Moussa, 
                   J. Rodr\'iguez-Carvajal, L. Pinsard, 
                   A. Revcolevschi, Y. M. Mukovskii, and D. Shulyatev,
                   \prb {\bf 64}, 104421 (2001);
                F.~Moussa, M. Hennion, F. Wang, P. Kober, 
                   J. Rodr\'iguez-Carvajal, P. Reutler, L. Pinsard, 
                   and A. Revcolevschi, 
                   {\it ibid.\/} {\bf 67}, 214430 (2003). 

\bibitem{End99} Y. Endoh, K. Hirota, S. Ishihara, S. Okamoto, 
                   Y. Murakami, A. Nishizawa, T. Fukuda, H. Kimura, 
                   H. Nojiri, K. Kaneko, and S. Maekawa,
                   \prl {\bf 82}, 4328 (1999).
                   
\bibitem{Fes01} A. Wei\ss{}e, J. Loos, and H. Fehske,
                   \prb {\bf 64}, 054406 (2001);
                   {\it ibid.\/} {\bf 68}, 024402 (2003).

\bibitem{Bri00} J. van den Brink, P. Horsch, and A. M. Ole\'s,
                   \prl {\bf 85}, 5174 (2000);
                W.-G. Yin, H.-Q. Lin, and C.-D.~Gong,
                   {\it ibid.\/} {\bf 87}, 047204 (2001).

\bibitem{Bal02} J. Ba\l{}a, A. M. Ole\'s, and P. Horsch,
                   \prb {\bf 65}, 134420 (2002).

\bibitem{Ram93} A. Ram\v{s}ak and P. Horsch,
                   \prb {\bf 48}, 10559 (1993);
                   {\it ibid.\/} {\bf 57}, 4308 (1998).

\bibitem{Bon92} J. Bon\v{c}a, P. Prelov\v{s}ek, I. Sega, H. Q. Lin,
                   and D.~K.~Campbell,
                   \prl {\bf 69}, 526 (1992).
                   
\bibitem{Wro96} P. Wr\'obel and R. Eder,
                   \prb {\bf 54}, 15882 (1996).

\bibitem{Ben99} P. Benedetti and R. Zeyher,
                   \prb {\bf 59}, 9923 (1999).

\bibitem{Hot00} T. Hotta, A. L. Malvezzi, and E. Dagotto,
                   \prb {\bf 62}, 9432 (2000).

\bibitem{Kil99} R. Kilian and G. Khaliullin,
                   \prb {\bf 60}, 13459 (1999).

\bibitem{Cha77} K. A. Chao, J. Spa\l{}ek, and A. M. Ole\'s,
                   J. Phys. C {\bf 10}, L271 (1977);
                   \prb {\bf 18}, 3453 (1978). 

\bibitem{noteaf} The FM ground state is found for realistic parameters 
                 of manganites. However, a somewhat smaller Hund's 
                 exchange $J_H$, or large AF superexchange $J'$ between 
                 $t_{2g}$ core spins would give instead an AF state
                 with uniformly polarized $|z\rangle$ orbitals at $n=1$.     

\bibitem{Hot03} T. Hotta,
                   \prb {\bf 67}, 104428 (2003).

\bibitem{Mil96} A. Millis,
                   \prb {\bf 53}, 8434 (1996).

\bibitem{Bal00} J. Ba\l{}a and A. M. Ole\'s,
                   \prb {\bf 62}, R6085 (2000).

\bibitem{Mar91} G. Mart\'inez and P. Horsch,
                   \prb {\bf 44}, 317 (1991).

\bibitem{Zac98} M. G. Zacher, E. Arrigoni, W. Hanke, 
                   and J. R. Schrieffer,
                   \prb {\bf 57}, 6370 (1998).
  
\bibitem{Sen00} D. S\'en\'echal, D. Perez, and M. Pioro-Ladri\`ere,
                   \prl {\bf 82}, 522 (2000).

\bibitem{Aic03} M. Aichhorn, M. Daghofer, H. G. Evertz, 
                   and W. von der Linden, 
                   \prb {\bf 67}, 161103(R) (2003).
    
\bibitem{Zaa88} J. Zaanen, A. T. Paxton, O. Jepsen, and O. K. Andersen, 
                   \prl {\bf 60}, 2685 (1988).

\bibitem{End97} Y. Endoh and K. Hirota,
                   J. Phys. Soc. Jpn. {\bf 66}, 2264 (1997).

\bibitem{Bac98} J. A. Fernandez-Baca, P. Dai, H. Y. Hwang, C. Kloc, 
                   and S-W. Cheong,                 
                   \prl {\bf 80}, 4012 (1998).

\bibitem{Mae03} S. Maezono and N. Nagaosa,            
                   \prb {\bf 67}, 064413 (2003).

\end{thebibliography}
\end{document}